\documentclass[final,5p,times,twocolumn]{elsarticle}
\usepackage{amssymb}
\usepackage{amsmath}
\usepackage{array}
\usepackage{makecell}
\usepackage{rotating}
\usepackage{booktabs}
\usepackage{graphicx}
\usepackage{lscape}
\usepackage{multirow}
\usepackage{geometry}
\usepackage[table]{xcolor}
\usepackage{float}
\usepackage{tabularx}
\usepackage{empheq}
\usepackage{soul}
\renewcommand{\hl}[1]{#1}
\usepackage[utf8]{inputenc}
\usepackage{newunicodechar}
\newunicodechar{₂}{$_2$}
\usepackage[font=normalsize,labelfont=bf]{caption}

\usepackage{nomencl}
\makenomenclature
\usepackage{array}
\usepackage{subfig}
\usepackage{color}
\newcolumntype{M}[1]{>{\centering\arraybackslash}m{#1}}
\usepackage{multirow}

\journal{Energy Conversion and Management}

\begin{document}

\begin{frontmatter}

\title{\hl{Ensuring reliability in 100\% renewable microgrids: a scenario-based
joint planning and operational design framework}}

\author[inst1]{Mohammed Zeehan Saleheen}
\ead{mohammed.saleheen@monash.edu}
\author[inst1,inst2]{Markus Wagner}
\ead{markus.wagner@monash.edu}
\author[inst1,inst2]{Hao Wang\corref{cor1}}
\cortext[cor1]{Corresponding author: Hao Wang.}
\ead{hao.wang2@monash.edu}

\affiliation[inst1]{organization={Department of Data Science \& AI, Faculty of IT, Monash University}, 
            addressline={Wellington Rd, Clayton}, 
            city={Melbourne},
            postcode={3800}, 
            state={VIC},
            country={Australia}}
\affiliation[inst2]{organization={Monash Energy Institute, Monash University}, 
            addressline={Wellington Rd, Clayton}, 
            city={Melbourne},
            postcode={3800}, 
            state={VIC},
            country={Australia}}

\begin{abstract}
Off-grid microgrids powered entirely by renewable energy sources face substantial challenges in achieving utility-grade reliability standards owing to renewable intermittency and the absence of grid backup. Existing microgrid planning frameworks often prioritize cost minimization while treating reliability as a secondary metric, thereby leading to suboptimal designs that are vulnerable to renewable variability, component failures, and other operational uncertainties. This paper presents a comprehensive scenario-based optimization framework that simultaneously addresses long-term capacity planning and short-term operational dispatch in two stages for 100\%-renewable microgrids. The developed two-stage stochastic programming model co-optimizes the investment and operation of photovoltaic generation and battery energy storage, while ensuring compliance with stringent reliability constraints following utility grid standards. Network modeling with operational constraints, such as line capacities and voltage limits, is incorporated to allow distributed resource placement leveraging power sharing between microgrid nodes. A novel scenario generation approach captures critical uncertainties, including seasonal demand fluctuations, solar output variations, and probabilistic equipment failures, through the statistical clustering of historical data. The optimization framework integrates utility-grade reliability constraints limiting the expected energy not served to below 0.002\% of the annual demand while minimizing the total system costs. Numerical simulations demonstrate the effectiveness of the proposed framework, achieving approximately 99.998\% supply reliability using only photovoltaic power and battery energy storage. The optimized network-aware distributed resource allocation provides inherent resilience through power rerouting during component outages, maintaining load continuity even under simultaneous equipment failures. This study confirms the feasibility of 100\%-renewable microgrids to support remote communities while meeting utility-grade reliability benchmarks.
\end{abstract}



\end{frontmatter}


\section*{Nomenclature}

\noindent\textbf{Acronyms and Abbreviations}
\begin{description}
    \item[PV] Photovoltaic generation unit.
    \item[BESS] Battery Energy Storage System.
    \item[ENS] Energy Not Served.
    \item[SOC] State of Charge.
    \item[O\&M] Operation and Maintenance.
    \item[VOLL] Value of Lost Load.
    \item[MTTR] Mean Time To Repair.
    \item[FTA] Fault Tree Analysis.
    \item[PCA] Principal Component Analysis.
    \item[NPV] Net Present Value.
    \item[C-rate] Battery charge/discharge rating.
\end{description}

\noindent\textbf{Sets}
\begin{description}
    \item[$N$] Set of nodes in the microgrid.
    \item[$E$] Set of distribution lines $(i,j)$.
    \item[$T$] Set of hourly time periods in the representative day.
    \item[$S$] Set of scenarios.
\end{description}

\noindent\textbf{Indices}
\begin{description}
    \item[$i,j$] Node indices, $i,j \in N$.
    \item[$t$] Time period index, $t \in T$.
    \item[$s$] Scenario index, $s \in S$.
\end{description}

\noindent\textbf{Parameters}
\begin{description}
    \item[$d_{i,t,s}$] Demand at node $i$ at time $t$ in scenario $s$.
    \item[$\eta_{\mathrm{dis}}$] Battery discharge efficiency.
    \item[$\eta_{\mathrm{ch}}$] Battery charge efficiency.
    \item[$\Delta t$] Duration of a time step (hour).
    \item[$soc^{\mathrm{init}}_{b,i}$] Initial state of charge of the battery at node $i$.
    \item[$soc_{\min}$] Minimum permissible state-of-charge fraction.
    \item[$soc_{\max}$] Maximum permissible state-of-charge fraction.
    \item[$P^{\max}_{ij}$] Thermal/ampacity limit of distribution line $(i,j)$.
    \item[$V_{\min}$] Minimum allowable voltage magnitude.
    \item[$V_{\max}$] Maximum allowable voltage magnitude.
    \item[$r$] Discount rate.
\end{description}

\noindent\textbf{Scenario-dependent parameters}
\begin{description}
    \item[$x^{\mathrm{fail}}_{b,i,s}$] Battery failure indicator at node $i$ in scenario $s$ (1 = failed, 0 = available).
    \item[$x^{\mathrm{fail}}_{pv,i,s}$] PV failure indicator at node $i$ in scenario $s$ (1 = failed, 0 = available).
\end{description}

\noindent\textbf{Planning (Stage 1) variables}
\begin{description}
    \item[$\mathrm{cap}^{pv}_{i}$] Installed PV capacity at node $i$.
    \item[$\mathrm{cap}^{b}_{i}$] Installed battery energy capacity at node $i$.
\end{description}

\noindent\textbf{Operational (Stage 2) variables}
\begin{description}
    \item[$p_{i,t,s}$] PV power output at node $i$ and time $t$.
    \item[$p^{b}_{i,t,s}$] Battery power at node $i$ and time $t$ 
          ($p^{b} > 0$ discharge, $p^{b} < 0$ charge).
    \item[$soc_{b,i,t,s}$] State of charge of battery at node $i$ and time $t$.
    \item[$P_{ij,t,s}$] Real power flow on line $(i,j)$ at time $t$.
    \item[$P_{ji,t,s}$] Reverse real power flow on line $(j,i)$.
    \item[$ENS_{i,t,s}$] Energy not served at node $i$ and time $t$.
\end{description}

\newpage

\section{Introduction}
\label{sec1}
Remote and rural communities worldwide often lack access to reliable electricity from national grids~\cite{akinyele2016strategy}. Extending centralized grid infrastructure to sparsely populated or hard-to-reach areas is substantially expensive~\cite{domenech2018local}, leaving millions with limited or no access to electricity~\cite{zebra2021review}. The achievement of universal electrification and development goals, e.g., Sustainable Development Goal 7 on affordable and clean energy, has prompted socioeconomic and policy support for standalone microgrid solutions.\hl{ For instance, Isihak et al.~\cite{isihak2023achieving} demonstrated that national rural electrification targets aligned with SDG7 can be cost-effectively met through optimized combinations of grid extension and standalone off-grid systems. Valencia-Díaz et al.~\cite{valencia2025optimal} further highlighted the role of standalone microgrids in sustainable remote development by jointly optimizing energy, water, and carbon objectives in isolated hybrid AC/DC configurations.  With such localized isolated networks equipped with their own generation and storage systems, standalone microgrids have emerged as a viable solution to electrify remote communities, offering a path to improved living standards and economic opportunity.} However, to ensure that these island microgrids deliver electricity with at least utility-grade reliability and affordability, they must be designed and operated in a reliable and cost-effective manner tailored to off-grid conditions.

In Australia, the standard of reliability at the grid level for supplying electricity is established by the Australian Energy Market Commission (AEMC), which mandates that the expected energy not served (EENS) should not exceed 0.002\% of the total annual demand~\cite{AEMC2024}. In grid-connected settings, this reliability standard is typically achieved with the support of the upstream grid, allowing microgrids to island during local faults and maintain supply through local generation until grid supply is restored. However, achieving such a stringent reliability standard in 100\% renewable-based off-grid settings is inherently challenging, due to the absence of grid support and the inherent variability and vulnerability of renewable resources. Historically, to meet a minimum level of reliability in off-grid settings, rural microgrids have been heavily relying on diesel generators. \hl{Sanni et al.~\cite{sanni2021analysis} illustrated this dependence by analyzing hybrid solar PV/diesel/biogas configurations designed specifically to provide backup power under unreliable supply conditions. Irfan et al.~\cite{irfan2024optimizing} similarly demonstrated the continued role of conventional generation in maintaining frequency stability within Australian microgrids.} However, while effective in ensuring supply continuity, this approach involves high fuel costs, complex logistics, and significant carbon emissions. \hl{Ranaboldo et al.~\cite{ranaboldo2015off} documented these limitations in off-grid community projects in Nicaragua, where high and variable fuel costs and the continuous requirement for fuel transportation to rural areas, contributed significantly to lifecycle expenditure. Furthermore, Rangel et al.~\cite{rangel2023optimisation} highlighted diesel generators in off-grid systems as a major source of CO₂ emissions and advocated limiting their use through renewable integration.}

In response, the falling costs of solar photovoltaics (PV) and battery energy storage systems (BESS)~\cite{tsianikas2019economic}, coupled with growing efforts to mitigate climate change, have spurred increasing interest in 100\%-renewable microgrid configurations. However, despite this technological potential and declining costs, the uptake of 100\% renewable-based microgrids in Australia remains limited. According to Wright et al.~\cite{wright2022australian} only a handful of such microgrid projects have moved beyond the pilot stage, indicating a gap between technological potential and practical application. This is because the economic viability and technical performance of 100\%-renewable microgrids depend on how well it balances cost and reliability. The achievement of this balance remains a complex multidimensional challenge that intersects technical feasibility~\cite{notton2018intermittent}, economic viability~\cite{saddari2025techno}, equipment adequacy~\cite{parag2019sustainable}, and increased operational complexity~\cite{rochd2025swarm}. This complexity is further amplified by the inherent variability and uncertainty associated with renewable energy resources, component failures, and evolving demand patterns. Therefore, a holistic, reliability-oriented planning framework is urgently needed to simultaneously address these challenges and support the cost-effective deployment of 100\% renewable microgrids.

The planning framework for microgrid design involves selecting an optimal combination of generation and storage capacities to meet local demand at minimal cost, while adhering to resource availability and reliability requirements. For example, \hl{Prathapaneni et al.~\cite{prathapaneni2019integrated} proposed an integrated framework for islanded microgrid planning that couples optimal capacity sizing with operational scheduling under uncertainty and further incorporates demand-side management. In a related but broader network context,  Wang et al.~\cite{wang2019joint} developed a stochastic joint planning model for active distribution networks that simultaneously determines renewable generation, battery storage, electric vehicle charging infrastructure, and network expansion decisions under renewable uncertainty. Huo et al.~\cite{huo2022reliability} then shifted attention toward reliability-aware storage planning by proposing a chance-constrained battery sizing method for island microgrids that explicitly links battery investment decisions with reliability performance. By contrast, Zhang et al.~\cite{zhang2017battery} addressed battery sizing from a grid-connected microgrid perspective, developing a method that simultaneously optimizes generation capacity and rule-based operational strategies in order to improve economic performance. At a more detailed operational level, Xie et al.~\cite{xie2021optimal} proposed a two-layer sizing strategy for smart microgrids with high photovoltaic penetration, incorporating virtual energy storage systems to enhance dispatch flexibility and reduce operational cost. Taken together, these studies represent substantial progress in individual aspects of microgrid planning, yet critical gaps persist in achieving a holistic approach that integrates reliability considerations, accounts for operational uncertainties, and reflects realistic network representation.}

\hl{The predominant paradigm in microgrid optimization has been largely shaped by cost minimization objectives, most commonly expressed through net present cost (NPC) or levelized cost of energy (LCOE) formulations. For example, Vaka and Matam~\cite{vaka2021optimal} focused on reducing operating costs through battery storage management strategies, while Kassab et al.~\cite{kassab2024optimal} extended this perspective by incorporating carbon emissions into a joint net present cost minimization framework. Similarly, Hafez et al.~\cite{hafez2021optimal} adopted hybrid metaheuristic method for economically efficient off grid system design, an approach echoed by Aeggegn et al.~\cite{aeggegn2024optimal}, who applied grey wolf optimization to achieve cost optimal sizing of multiple interconnected microgrids. Such economically focused paradigms, while incorporating capacity adequacy constraints to maintain basic supply feasibility, often fail to ensure reliable system operation under real world conditions. This limitation is further reflected in studies that attempt to address uncertainty, yet remain anchored within the same cost driven orientation.

Yang et al.~\cite{yang2023chance}, for instance, addressed supply and demand uncertainty through chance constrained dispatching, while Mokhtara et al.~\cite{mokhtara2020integrated} considered integrated supply and demand management for off grid systems in arid environments. Copp et al.~\cite{copp2022optimal} further illustrate this tendency through an optimization framework that determines the minimum renewable generation and storage capacities required to meet demand over multiyear horizons, yet remains primarily oriented toward resource adequacy and cost minimization.} Similarly, Jeyaprabha et al.~\cite{jeyaprabha2022probabilistic} and Bilal et al.~\cite{bilal2025hybrid} adopt comparable cost minimization frameworks, implicitly assuming that reliability can be achieved through conservative over-sizing of renewable and storage capacities. However, across these studies, the resulting formulations primarily ensure that capacity installations remain feasible or non negligible, with reliability treated as an implicit outcome rather than a rigorously enforced design criterion.

In addition, cost-centric approaches often fail to deliver the promised performance under real-world operational conditions due to their inability to explicitly incorporate reliability considerations. These methodologies exhibit several critical shortcomings: (i) the absence of quantitative reliability targets results in either overly conservative designs with excessive costs or underperforming systems that fail to meet reliability expectations; (ii) the lack of explicit reliability constraints prevents the optimization from exploring the cost-reliability trade-off space effectively; and (iii) the assumption that oversizing automatically enhances reliability overlooks the complex interactions among component failures, renewable variability, and system-level performance.

Recognizing the limitations of purely economic objectives, recent research has increasingly sought to incorporate reliability considerations into microgrid planning through multi-objective formulations and penalty-based approaches. For example, Wang et al.~\cite{wang2021three} introduced a multi-objective framework that explicitly balances cost minimization with resilience maximization in networked microgrids, particularly targeting extreme outage scenarios, marking a conceptual shift toward the explicit treatment of reliability. In parallel, several studies have explored penalty-based methods that embed reliability within the economic objective. Rasool et al.~\cite{rasool2022scenario} proposed a techno-economic reliability framework that jointly minimizes the levelized cost of energy (LCOE) and unserved energy, effectively quantifying reliability as a cost component. Wu et al.~\cite{wu2021milp} again extended this paradigm by incorporating expected load curtailment costs along with capital expenditures. More recently, Nallainathan et al.~\cite{sakthivelnathan2024cost} refined this approach by combining investment costs with EENS penalties to choose a cost-effective reliability level in 100\% renewable microgrid planning.

Although multiobjective and penalty-based approaches represent meaningful progress toward incorporating reliability into microgrid planning, they often do not guarantee specific reliability targets unless penalty coefficients are carefully calibrated, which is often done arbitrarily. In addition, treating reliability as a soft constraint allows optimization to compromise reliability under cost pressure. Furthermore, beyond these limitations, most existing studies continue to focus primarily on resource adequacy, evaluating reliability based on the availability of renewable generation and storage capacity. \hl{This tendency is evident in the work of Chebabhi et al.~\cite{chebabhi2023optimal}, who size renewable microgrids primarily based on financial considerations, implicitly treating reliability as a byproduct of sufficient capacity margins. Similarly, Mathew et al.~\cite{mathew2022sizing}, in a comprehensive review, observe that the majority of microgrid sizing approaches follow this paradigm, relying on resource availability rather than explicitly defined reliability metrics.} However, these studies often overlook the probabilistic nature of component failures or assume perfect component availability. These assumptions contradict the real-world operational experience, where component failures are the main contributor to system unreliability. In addition, resource adequacy models often fail to capture the cascading effect of component failures on system performance, which is crucial in investigating system operation. 

Another pervasive practice in the literature involves treating reliability as a post-design evaluation metric rather than integrating it into the core planning framework. \hl{Ahshan et al.~\cite{ahshan2017microgrid} followed this approach by first designing a microgrid and then separately evaluating its reliability under renewable intermittency conditions. Nargeszar et al.~\cite{nargeszar2023reliability} adopted a similar decoupled workflow, assessing the reliability of renewable-based microgrids only after the capacity decisions had been finalized. Although such post-hoc evaluations provide insight into system performance, they inherently separate the optimization process from reliability considerations.} In addition, such approach often prevents designers from exploring the interdependencies between investment decisions and reliability outcomes. Moreover, when a design fails to meet reliability standards during evaluation, it necessitates iterative redesign processes that are computationally expensive and may not converge to optimal solutions.

\hl{Interestingly, a large proportion of microgrid design studies adopt single node or aggregated system representations, wherein generation, storage, and demand are collectively modeled as a unified entity. For example, Krishna et al.~\cite{krishna2024robust} sized a renewable rich multi source microgrid using a fully aggregated system model, while Nurunnabi et al.~\cite{nurunnabi2019size} conducted sensitivity analysis of a hybrid wind and photovoltaic microgrid within a single bus framework. Similarly, Shezan et al.~\cite{shezan2023optimization} optimized an islanded hybrid microgrid without explicit network disaggregation, and Firdouse and Surender Reddy~\cite{firdouse2023hybrid} designed hybrid energy storage systems under the same simplifying assumption. This modeling paradigm extends beyond isolated instances and reflects a broader methodological convention in the field.} For example, Copp et al.~\cite{copp2022optimal} and Rasool et al.~\cite{rasool2022scenario} both employ aggregated demand node models that represent entire service areas as single points without accounting for internal network topology. Similarly, Wu et al.~\cite{wu2021milp} optimized generation and storage capacities for standalone microgrids under centralized system assumptions that disregard power routing complexities and node specific reliability variations. 

While such simplified representations offer computational tractability, they fundamentally misrepresent the distributed nature of microgrids and obscure critical network level phenomena that bear directly on system performance. In particular, the treatment of reliability in aggregated network model remains a critical shortcoming, as reliability is assessed globally rather than at the level of individual nodes or network segments. Hence, this centralized representation inherently ignores the potential reliability benefits of distributed generation, including supply diversity and localized load support during contingency events. Furthermore, the absence of explicit network topology precludes the assessment of line capacity limits and power flow constraints, raising serious concerns about whether designs derived under such assumptions remain implementable in physical networks where these constraints are binding.

Based on the above analysis, the existing literature still lacks a holistic and reliability-oriented planning framework for the design of a 100\% renewable-based microgrid. Key gaps include the absence of explicit reliability constraints within the optimization process when considering off-grid settings. Rather than enforcing predefined reliability standards, such as utility-grade reliability targets mandated by regulatory bodies, most studies rely on soft penalty terms for unserved energy or perform post-solution reliability evaluations. Moreover, equipment failure modeling is often ignored, with reliability assessed under the assumption of perfect component availability, a simplification that can significantly overestimate system performance. Furthermore, almost all models rely on centralized or single-node representations, ignoring network topology, line constraints, and node-level reliability variations. As a result, critical factors such as power flow limitations, node-specific investment, and failures are often excluded from the analysis. Together, these limitations highlight the need for an integrated planning approach for 100\% renewable-based microgrids that incorporates reliability considerations, including equipment failure, network constraints, and reliability metrics, directly within the core optimization process. Table 1 provides a comparative analysis of the features of the new model proposed in this article and those considered in the literature. The main contributions of this study are as follows.

\begin{itemize}
    \item \textit{Reliability-Oriented Joint Optimization Framework:} This work introduces a reliability-oriented microgrid planning framework explicitly integrating utility-grade reliability targets into joint planning and operational optimization for 100\%-renewable microgrids. Unlike conventional cost-centric or post hoc approaches, our design framework guarantees compliance with reliability regulatory benchmarks, providing reliable microgrid designs suitable for off-grid electrification.
    \item \textit{Topology-Aware Distributed Resource Investment and Operation:} \hl{The developed model explicitly incorporates network topology and constraints, enabling node-specific investment decisions. This contrasts with the centralized single-node models predominant in the literature (Table 1) and reveals that optimal reliability--constrained allocation produces a strategically asymmetric design: universal battery deployment for local reliability buffering combined with selective PV placement for network-level generation sharing.}
    \item \textit{Uncertainty-Aware Scenario Generation Methodology:} An integrated probabilistic scenario generation methodology is proposed, combining statistical clustering for renewable and demand uncertainties with Fault Tree Analysis (FTA) for equipment failure modeling, to ensure robust and reliable operation.
    \item \textit{Simulation-driven Planning Insights:} Numerical simulations on the IEEE 33-bus islanded microgrid demonstrate that the proposed framework achieves approximately 99.998\% reliability using only photovoltaic generation and battery storage. The results reveal that the optimized distributed resource placement provides inherent resilience through network-level power sharing, enabling the microgrid to maintain uninterrupted supply even under simultaneous component failures without requiring local backup at every node. These findings confirm the practical feasibility of 100\%-renewable microgrids that meet utility-grade reliability benchmarks while avoiding over-investment through topology-aware design.
\end{itemize}

The remainder of the paper is organized as follows. \hl{Sections 2–4 collectively present the proposed methodology. Section 2 defines the microgrid system model and the two-stage optimization structure that separates long-term planning decisions from short-term operational dispatch.} Section 3 details the scenario generation methodology, describing the statistical approach to capture uncertainties in load, PV generation, and component failures. Section 4 presents the mathematical formulation of the optimization framework, including the objective function and the constraints that enforce reliability and techno-economic feasibility. Section 5 discusses the validation results using the IEEE 33-bus test system, analyzing the performance and reliability of the optimized microgrid design. Finally, Section 6 concludes the study, summarizing key findings and suggesting future research directions.

\begin{sidewaystable*}[htbp]
\caption{Comparison of existing studies with the proposed work.}
\centering
\scriptsize
\renewcommand{\arraystretch}{1.5} 
\setlength{\tabcolsep}{3pt} 

\begin{tabular}{
>{\centering\arraybackslash}p{1.5cm}
>{\centering\arraybackslash}p{1.7cm}
>{\centering\arraybackslash}p{3.4cm}
>{\centering\arraybackslash}p{1.7cm}
>{\centering\arraybackslash}p{2.2cm}
>{\centering\arraybackslash}p{2.3cm}
>{\centering\arraybackslash}p{2.4cm}
>{\centering\arraybackslash}p{1.9cm}
>{\centering\arraybackslash}p{3.0cm}
>{\centering\arraybackslash}p{1.8cm}
}
\toprule
\textbf{Study} &
\textbf{\makecell{100\%\\Renewable}} &
\textbf{\makecell{Reliability\\Consideration}} &
\textbf{\makecell{Equipment\\Failure}} &
\textbf{\makecell{Node-specific\\Investment}} &
\textbf{\makecell{Scenario-based\\Modeling}} &
\textbf{\makecell{Two-Stage\\Optimization}} &
\textbf{\makecell{Component \\Deployment}} &
\textbf{Objective} &
\textbf{\makecell{PF\\Constraints}} \\
\midrule
Copp et al. (2022)~\cite{copp2022optimal} & \makecell{Yes} & \makecell{No} & \makecell{No} & \makecell{No} & \makecell{No} & \makecell{No} & \makecell{Centralized} & \makecell{LCOE} & \makecell{No} \\
Kamal et al. (2022)~\cite{kamal2022planning} & \makecell{No} & \makecell{Post-evaluation} & \makecell{No} & \makecell{No} & \makecell{No} & \makecell{No} & \makecell{Centralized} & \makecell{Investment} & \makecell{No} \\
Nallainathan et al. (2023)~\cite{sakthivelnathan2024cost} & \makecell{Yes} & \makecell{Post-evaluation} & \makecell{Partial} & \makecell{No} & \makecell{Partial} & \makecell{Yes} & \makecell{Centralized} & \makecell{Investment\\+ EENS} & \makecell{No} \\
Rasool et al. (2022)~\cite{rasool2022scenario} & \makecell{Yes} & \makecell{Post-evaluation} & \makecell{No} & \makecell{No} & \makecell{Yes} & \makecell{Partial} & \makecell{Centralized} & \makecell{LCOE + UEL} & \makecell{No} \\
Wu et al. (2021)~\cite{wu2021milp} & \makecell{No} & \makecell{Yes} & \makecell{No} & \makecell{No} & \makecell{No} & \makecell{No} & \makecell{Centralized} & \makecell{Investment + \\ Curtailment Cost} & \makecell{No} \\
Wang et al. (2021)~\cite{wang2021three} & \makecell{No} & \makecell{No} & \makecell{Yes} & \makecell{No} & \makecell{No} & \makecell{Yes} & \makecell{N/A} & \makecell{Resilience} & \makecell{Yes} \\
Jeyaprabha et al. (2023)~\cite{jeyaprabha2022probabilistic} & \makecell{No} & \makecell{No} & \makecell{No} & \makecell{No} & \makecell{No} & \makecell{No} & \makecell{Centralized} & \makecell{LCOE} & \makecell{No} \\
Bilal et al. (2024)~\cite{bilal2025hybrid} & \makecell{No} & \makecell{No} & \makecell{No} & \makecell{No} & \makecell{No} & \makecell{No} & \makecell{Centralized} & \makecell{LCOE} & \makecell{No} \\
Krishna P.M et al. (2024)~\cite{krishna2024robust} & \makecell{Yes} & \makecell{No} & \makecell{No} & \makecell{No} & \makecell{No} & \makecell{No} & \makecell{Centralized} & \makecell{LCOE} & \makecell{No} \\
Camargo et al. (2024)~\cite{krishna2024robust} & \makecell{No} & \makecell{N-1 reserve} & \makecell{No} & \makecell{No} & \makecell{No} & \makecell{No} & \makecell{Centralized} & \makecell{Investment +\\ Operational} & \makecell{Yes} \\
\midrule
\textbf{This Work (2025)} &
\makecell{Yes} &
\makecell[t]{Explicit reliability \\ constrained  + \\ Utility-grade reliability \\ assurance} &
\makecell{Yes} &
\makecell{Yes} &
\makecell{Yes} &
\makecell{Yes} &
\makecell{Distributed} &
\makecell[t]{Investment + \\ Operational + \\ Reliability Penalty} &
\makecell{Yes} \\
\bottomrule
\end{tabular}
\label{tab:comparison}
\end{sidewaystable*}

\section{Problem Overview}

\hl{This section introduces the system model and the two-stage optimization structure that underpins the proposed framework. The system model defines the microgrid topology, distributed energy resources, and the operational scenarios considered, while the two-stage structure formalizes the separation between long-term investment decisions and short-term operational dispatch.}

\subsection{\hl{System description and optimization structure}}

We assume a microgrid that is isolated from the main grid distribution network. It has PV generation units, battery energy storage systems, and a set of loads distributed across nodes N. The system operates under multiple scenarios $s \in \mathcal{S}$, which encapsulate key uncertainties, including stochastic demand fluctuations, intermittent photovoltaic (PV) generation, and probabilistic component failures.

\begin{figure}[!t] 
    \centering 
    \includegraphics[width=1\linewidth]{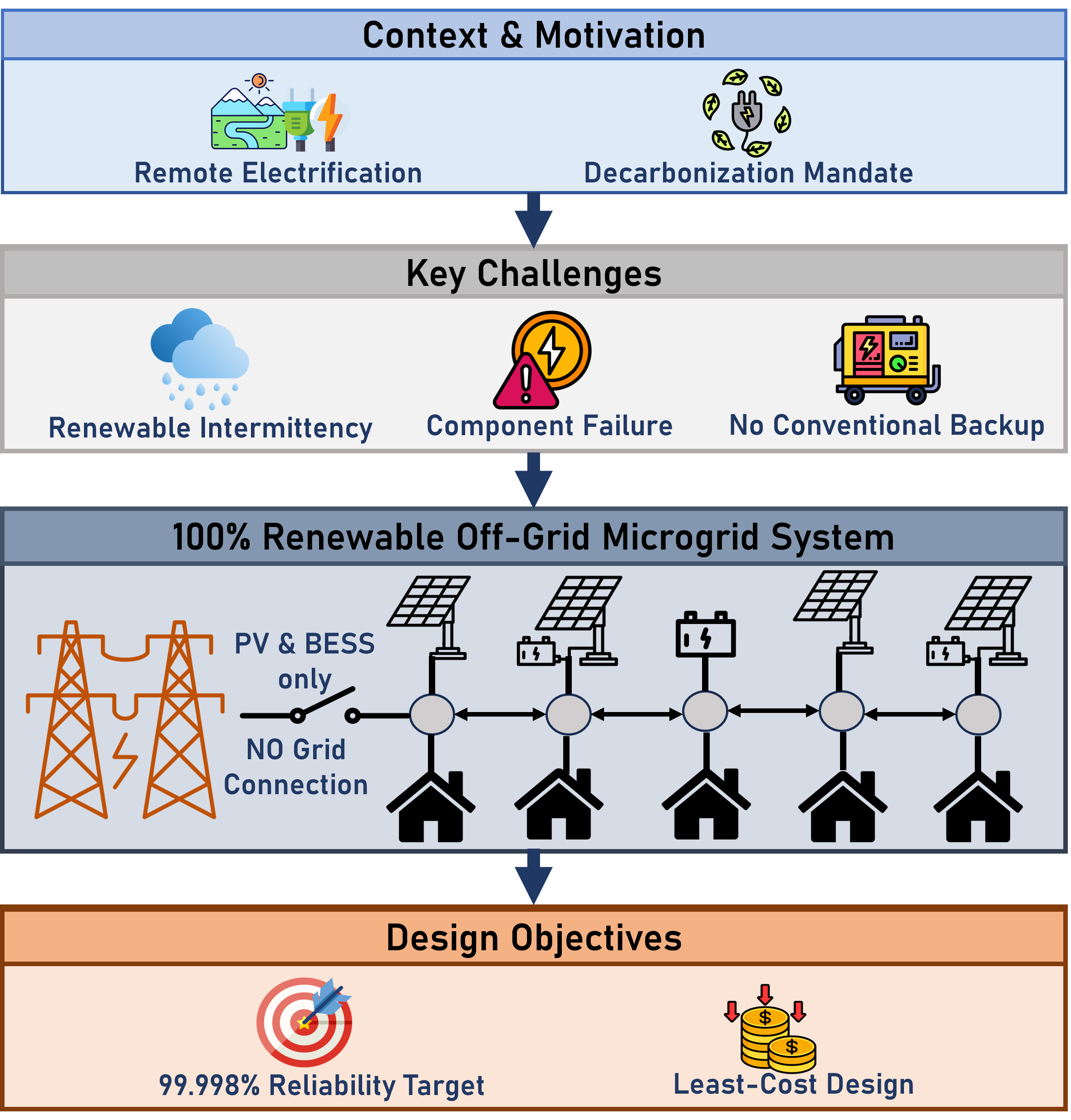} 
    \caption{\hl{Logical framework of a 100\% renewable-based off-grid microgrid design.}} 
    \label{Fig.1}  
\end{figure}

\begin{figure*}[!t] 
    \centering 
    \includegraphics[width=0.8\linewidth]{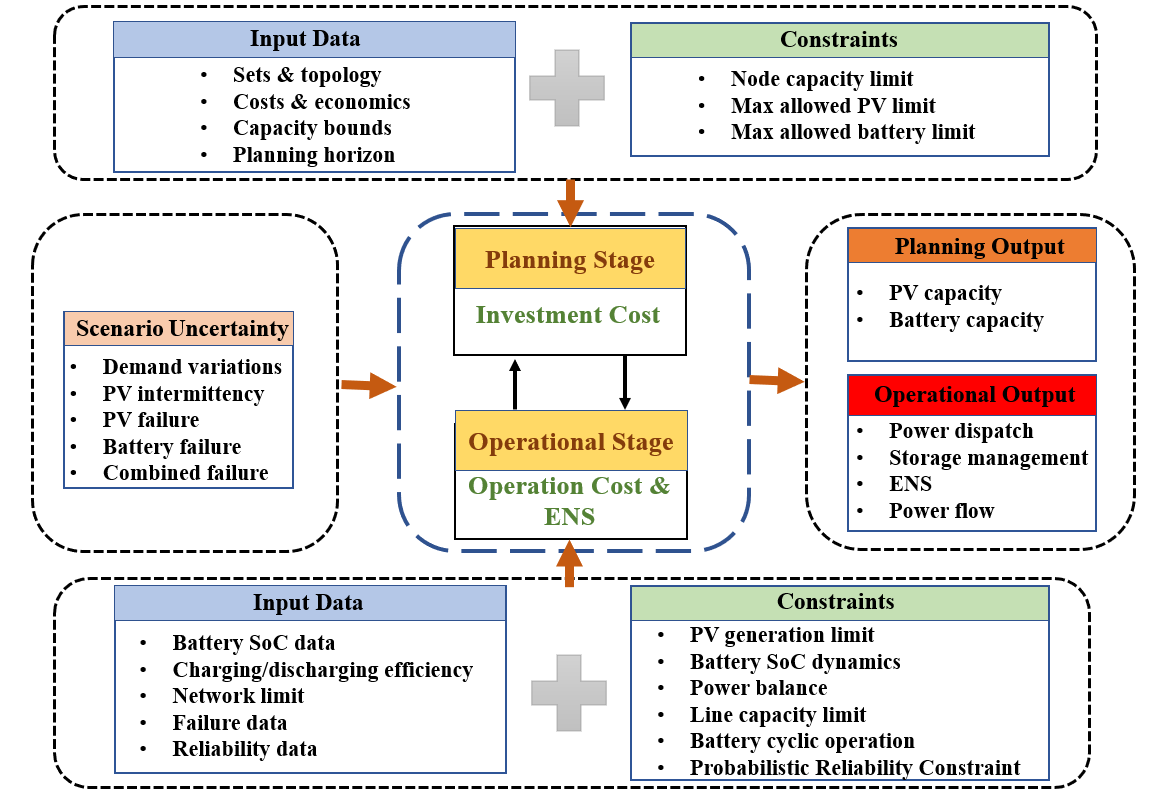} 
    \caption{Joint planning and operational framework for the design of 100\%-renewable microgrid.} 
    \label{Fig.2}  
\end{figure*}

The planning horizon $H$ spans 10 years, while the operational stage is represented through a set of 24-hour scenarios derived from clustered representative days. \hl{It is important to note that the representative days are modeled independently and do not preserve chronological links across days. As a result, inter-day energy transfer, including long-term storage dynamics, is not explicitly represented. Instead, each representative day is interpreted as a typical and repeatable operating pattern within the year, which is appropriate for long-term planning purposes.} Each representative day is associated with multiple operational scenarios that capture variability and failure conditions, with scenario weights reflecting their relative annual significance. The detailed scenario generation methodology has been provided in the upcoming section. Fig.~\ref{Fig.1} illustrates the logical framework of the 100\% renewable-based off-grid microgrid design, mapping design challenges to targets, while Fig.~\ref{Fig.2} presents an overview of the proposed reliability-oriented joint planning and operational framework. In Fig.~\ref{Fig.2}, the planning stage determines the optimal capacities of PV units and batteries, while the operational stage manages power dispatch and storage under uncertainties in demand, PV generation, and component failures.  To address these uncertainties, the model employs scenario based statistical techniques to evaluate the system reliability.\\

\subsection{Two-stage Structure}

We conceptualize the overall problem as a two-stage optimization model:
\begin{itemize}
    \item \textit{The Planning Stage (Stage 1)} determines long-term investment decisions, including the capacities of PV units ($\text{cap}_i^\text{pv}$)  and community batteries ($\text{cap}_i^\text{b}$) at each node $i \in \mathcal{N}$.
    \item \textit{The Operational Stage (Stage 2)} optimizes the operation of the installed resources in various scenarios that represent uncertainties in demand, generation, and component failures.
\end{itemize}

\hl{This proposed framework constitutes a classical two-stage stochastic decision structure. In Stage 1, investment decisions (PV and battery capacities) are determined ``\textit{here-and-now}'', prior to the realization of uncertainty. In Stage 2, operational recourse decisions (dispatch, storage scheduling, and power flows) adapt to each scenario in a ``\textit{wait-and-see}'' manner. Non-anticipativity of first-stage decisions is enforced structurally, as investment variables are defined independently of scenarios, while second-stage variables are scenario-specific. For computational tractability, the model is implemented and solved using its deterministic equivalent (extensive form), in which all scenarios are embedded within a single mixed-integer linear programming formulation.}\\

\section{Scenario Generation Methodology}

Real-world off-grid microgrids operate under multiple uncertainties that significantly affect both planning and operational decisions. \hl{The consequences of inadequately accounting for such uncertainties have been well documented in the literature. Mansouri et al.~\cite{mansouri2023using} demonstrated that neglecting load uncertainty alone can lead to substantially suboptimal microgrid operation plans, while Borujeni et al.~\cite{borujeni2017accurate} established that accurate characterization of uncertainty dynamics is a prerequisite for reliable planning outcomes rather than a mere analytical refinement.} Given these findings, the load demand in particular exhibits considerable variability due to daily and seasonal patterns~\cite{li2018seasonal}, unpredictable consumer behavior~\cite{zeng2025human}, and changing socio--economic conditions, creating temporal fluctuations in electricity consumption. On the other hand, PV generation also presents substantial intermittency challenges caused by weather variations~\cite{soliman2022numerical}, cloud cover, and site-specific geographical characteristics~\cite{tang2019evaluating}, leading to unpredictable power output patterns. Again, system reliability is often compromised by component failures, such as those in photovoltaic panels and battery storage systems, each exhibiting distinct failure characteristics and rates. The simultaneous occurrence of these above uncertainties creates a complex optimization environment that requires comprehensive scenario generation methodologies to capture both individual uncertainty patterns and their interactions, thereby ensuring robust microgrid designs that perform reliably under diverse real-world conditions.

To address this, a comprehensive statistical scenario generation framework is developed to systematically capture the stochastic characteristics of load demand variability, solar irradiance fluctuations, and component failure events. Component failures are incorporated by assigning annual failure rates of 0.5/year for PV units and 0.2/year for battery systems \cite{gholami2023enhanced}, with typical outage durations characterized by representative mean time to repair of 8 hours and 4 hours, respectively. In this research, the hourly load demand for a rural distribution network (SMR8) in Victoria, Australia is used from~\cite{TeamNando}, and the synthetic solar PV generation data (normalized output per unit capacity) is obtained from renewable.ninja~\cite{renewablesninja}. Moreover, we assume that this solar PV generation profile is uniform across all nodes due to their geographical proximity within the network. The scenario generation method is outlined below.

\subsection{Feature Construction}

In this stage, daily feature vectors are constructed to capture the 24-hour operational profile of the microgrid. Each day’s feature vector concatenates the hourly load profiles of all 33 nodes with the corresponding PV generation profile for that day. For each day $d$ in the year, a comprehensive feature vector is constructed by combining load and PV characteristics:
For each day (d), the load vector for node (i) is defined as:
\begin{equation}
\label{eq:feature_vector}
\begin{split}
    \mathbf{f}_d &= \left[ \mathbf{L}_{d,1}, \mathbf{L}_{d,2}, \dots, \mathbf{L}_{d,33}, \mathbf{P}_d \right], \\
    \mathbf{L}_{d,i} &= [L_{d,i}(1), \dots, L_{d,i}(24)], \\
    \mathbf{P}_d &= [P_d(1), \dots, P_d(24)],
\end{split}
\end{equation}
where $L_{d,i}(t)$ is the load at node $i$ during hour $t$ (1,...,24) of day $d$, and $P_d(t)$ is the normalized PV output at hour $t$ (1,...,24) of day $d$. In other words, we form a high-dimensional feature vector by appending the 24-hour load profile of each of the 33 nodes and the 24-hour PV profile. The PV generation profile is normalized to its peak capacity so that it is on a comparable scale with the load data. This comprehensive feature representation preserves the diurnal shape of demands and generation, allowing clustering algorithms to discern patterns across entire daily cycles. 

The daily feature matrix for the entire year becomes:
\begin{equation}
\mathbf{X} = \begin{bmatrix} \mathbf{f}_1 \\ \mathbf{f}_2 \\ \vdots \\ \mathbf{f}_{365} \end{bmatrix} \in \mathbb{R}^{365 \times D}, \quad D = (N \times T) + T, 
\label{eq:daily_feature_matrix}
\end{equation}
where 365 is the number of days in the year, and the dimension of the feature vector is $D$ = 816 ($N = 33$ is the number of nodes and $T$ is 24 hours, each day’s vector contains 33 × 24 = 792 load values and 24 PV values = 816 total).

\subsection{Seasonal Stratification}

To capture the inherent seasonal variations in both load demand and PV generation patterns, the annual dataset is stratified into four distinct seasons based on climatic conditions \cite{Season}. We denote the seasons as $s \in \{\text{Spring},\text{Summer}, \text{Autumn}, \text{Winter}\}$:
\begin{itemize}
\item \textit{Spring}: September, October, November;
    \item \textit{Summer}: December, January, February;
    \item \textit{Autumn}: March, April, May;
    \item \textit{Winter}: June, July, August.
\end{itemize}

For each season $s$, the corresponding subset of daily features is extracted as:
\begin{equation}
\mathbf{X}_s = \{ \mathbf{f}_d : d \in \mathcal{D}_s \},
\label{eq:seasonal_subset}
\end{equation}
where $\mathcal{D}_s$ represents the set of days belonging to season $s$.

\subsection{Feature Normalization and Dimensionality Reduction}

To ensure equitable treatment of different feature scales and to improve clustering performance, each seasonal dataset undergoes Min-Max normalization~\cite{han2012data}, which is applied column-wise to each feature:
\begin{equation}
\mathbf{X}_{s,\text{norm}} = \frac{\mathbf{X}_s - \min(\mathbf{X}_s)}{\max(\mathbf{X}_s) - \min(\mathbf{X}_s)}.
\label{eq:minmax}
\end{equation}

Given the high dimensionality of the feature space (816 dimensions), Principal Component Analysis (PCA)~\cite{rouani2021shading} is applied to reduce computational complexity while preserving essential information.
\begin{equation}
\mathbf{X}_{s,\text{pca}} = \mathbf{X}_{s,\text{norm}} \, \mathbf{W}_{\text{pca}},
\label{eq:pca}
\end{equation}
where $\mathbf{W}_{\text{pca}}$ contains the principal component loadings that retain 95\% of the total variance.

\subsection{Clustering Analysis}
For each season, the optimal number of clusters is determined using a two-stage approach that combines silhouette analysis with objective function convergence analysis. The silhouette coefficient~\cite{lengyel2019silhouette} for a given clustering configuration is calculated as:
\begin{equation}
S = \frac{1}{n} \sum_{i=1}^{n} \frac{b_i - a_i}{\max(a_i, b_i)},
\label{eq:silhouette}
\end{equation}
where $a_i$ is the average distance from point $i$ to other points in the same cluster, $b_i$ is the average distance from point $i$ to points in the nearest neighboring cluster, and $n$ is the total number of data points.

Initial analysis of silhouette scores across the full range $K \in [2, 20]$ revealed that pure silhouette maximization tends to favor low cluster numbers (typically $K = 2$–$3$) for each season. However, preliminary optimization tests demonstrated that such low cluster numbers fail to adequately capture the operational diversity required for robust microgrid design, leading to suboptimal solutions.

To address this limitation, a convergence analysis of the microgrid optimization objective function was conducted for varying cluster numbers. This analysis revealed that the objective function stabilizes when $K \geq 7$, indicating that at least seven representative days per season are necessary to maintain the quality of the solution.

Therefore, the optimal number of clusters $K^*$ for each season is selected as:
\begin{equation}
K^* = \arg\max_{K \in [7, 20]} S(K),
\label{eq:optimal_k}
\end{equation}
subject to the constraint  $K^* \geq 7$ to ensure an adequate representation of seasonal operational diversity. 

\subsection{Representative Day Selection}
K-means clustering is applied to each seasonal dataset using the optimal number of clusters. For each cluster $c$, the representative day is selected as the day closest to the cluster centroid in the PCA feature space:
\begin{equation}
d_{\text{rep},c} = \arg\min_{d \in C_c} \left\| \mathbf{x}_{d,\text{pca}} - \boldsymbol{\mu}_c \right\|_2,
\label{eq:rep_day_selection}
\end{equation}
where $C_c$ represents the set of days assigned to cluster $c$, $\boldsymbol{\mu}_c$ is the centroid of cluster $c$, and $\mathbf{x}_{d,\text{pca}}$ is the PCA-transformed feature vector for day $d$.

The weight assigned to each representative day is proportional to the size of its corresponding cluster:
\begin{equation}
w_{\text{rep},c} = \frac{|C_c|}{365}.
\label{eq:rep_day_weight}
\end{equation}

\subsection{Failure Scenario Integration}

To systematically incorporate component reliability into the planning formulation, we construct a set of three \emph{mutually exclusive} daily failure-state scenarios based on fault tree analysis (FTA)~\cite{jaise2013power}, using failure rates and mean times to repair (MTTR) for PV units and batteries. The scenario design is motivated by the observation that, for the two critical distributed energy resource (DER) classes considered in this study (PV units and batteries), the dominant daily reliability states can be represented parsimoniously as: \textbf{(i)} \emph{Normal operation}, in which all installed PV units and batteries remain fully available throughout the day; \textbf{(ii)} \emph{Single-component outages}, in which PV-only or battery-only outages occur at one or more nodes, and \textbf{(iii)} \emph{Combined outages}, in which coincident outages affect both PV and battery units at one or more nodes simultaneously. This three-scenario structure provides a principled yet compact representation of the failure-state space. Enumerating all possible node-level outage combinations would grow combinatorially with network size and quickly render the stochastic program intractable. Instead, the proposed decomposition aggregates failure events by class-level co-occurrence, capturing the qualitatively distinct operational regimes that primarily govern energy-not-supplied (ENS). The underlying rationale is that the system's ability to maintain supply adequacy differs fundamentally across these regimes. Under nominal operation, installed capacity is tested against resource variability alone. Single-class outages probe the system's ability to compensate for the loss of one resource type through the other. Concurrent outages, although least probable, stress-test the microgrid under compound failure conditions where both generation and storage are simultaneously degraded. By preserving these distinct reliability regimes within a computationally tractable scenario tree, the formulation strikes a deliberate balance between modeling fidelity and scalability, which is particularly important for planning-stage optimization problems. The annual failure rates, denoted \(\lambda_{\text{PV}}\) for PV units and \(\lambda_{\text{B}}\) for batteries, are transformed into per-day, per-bus probabilities under a Poisson process assumption, yielding
\begin{equation}
p_{\text{PV}}^{(d)} = 1 - e^{-\lambda_{\text{PV}}/365}, 
\qquad 
p_{\text{B}}^{(d)} = 1 - e^{-\lambda_{\text{B}}/365}.
\label{eq:poisson-day}
\end{equation}

We assume baseline independence between components and across buses \(i \in \mathcal{N}\). This assumption is a reasonable first-order approximation because PV units and batteries rely on distinct technologies with largely independent failure mechanisms (e.g., semiconductor degradation or inverter related faults for PV units versus electrochemical aging for batteries), and the nodes are geographically dispersed within the distribution network, making spatially correlated failures unlikely. \hl{However, we acknowledge that correlated failures can arise in practice under certain conditions, such as extreme weather events affecting multiple nodes simultaneously, shared balance-of-system components (e.g., common inverter platforms or control infrastructure), or systemic supply-chain defects across identical equipment. It is worth noting that the combined failure scenario considered in this study, where simultaneous PV and battery outages occur at the same node, partially captures the operational impact of correlated failures by representing the worst-case loss of both generation and storage resources at a given location. Future extensions of this framework could incorporate common-cause failure models or copula-based dependency structures to more rigorously capture spatial and technological correlations across nodes and components, thereby providing a more comprehensive characterization of correlated failure risks in islanded microgrids.} 

Under the baseline independence assumption, the probability of \emph{normal operation} (no outages anywhere in the network) is
\begin{equation}
P_{\text{normal}} 
= \prod_{i \in \mathcal{N}} \big(1 - p_{\text{PV}}^{(d)}\big)\big(1 - p_{\text{B}}^{(d)}\big).
\label{eq:Pnormal}
\end{equation}
The probability of a \emph{combined} outage—at least one bus experiencing simultaneous PV and BESS failure \((k \geq 1)\)—is
\begin{equation}
P_{\text{combined}} 
= 1 - \prod_{i \in \mathcal{N}} \Big(1 - p_{\text{PV}}^{(d)} \, p_{\text{B}}^{(d)}\Big),
\label{eq:Pcombined}
\end{equation}
while the \emph{single-component} failure probability (exclusive PV or BESS outage at one or more buses) constitutes the residual probability mass:
\begin{equation}
P_{\text{single}} = 1 - P_{\text{normal}} - P_{\text{combined}}.
\label{eq:Psingle}
\end{equation}
By construction, \(P_{\text{normal}} + P_{\text{single}} + P_{\text{combined}} = 1\). These probabilities serve as weights for the daily scenarios, with MTTR informing the sampling of outage durations within each scenario.

\begin{itemize}
    \item \textbf{Normal Operation Scenario} \\
    This scenario assumes the perfect operation of all components and occurs with the probability $P_{\text{normal}}$, as defined in Eq. (11). This represents the baseline operation for the representative day, characterized by full PV generation and battery availability. 

    \item \textbf{Single-Component Failure Scenario} \\
    This scenario considers the failure of either a PV unit or battery storage system with the probability $P_{\text{single}}$. The failure profile generation follows a stochastic process where:
    \begin{itemize}
        \item PV failure: PV generation experiences an outage on that day, while the battery operates normally.
        \item Battery failure: The battery energy storage system fails (becomes unavailable), while PV generation is still available.
        \item Failure timing: Uniformly distributed over the 24-hour period.
        \item Failure duration: Exponentially distributed with a mean equal to the component's Mean Time To Repair (MTTR).
    \end{itemize}

    \item \textbf{Combined Failure Scenario} \\
    This scenario represents the simultaneous failure of both PV and battery systems at the same node with the probability $P_{\text{combined}}$. This constitutes an extreme contingency. Although rare, it is included to ensure robustness in the reliability analysis. In this scenario, the affected node loses both local generation and storage support for a duration determined by the overlapping outage periods, relying solely on network support to meet loads.
\end{itemize}

\hl{It is important to note that, while the three failure categories define the class-level co-occurrence structure, the specific failure realization within each category is stochastic. For each representative day, the failing node(s) are randomly sampled, either a single node or multiple nodes. The failure start time is uniformly distributed across the 24-hour period, and the failure duration is drawn from an exponential distribution with mean equal to the component’s MTTR. Across the 30 representative days, these stochastic realizations collectively sample a diverse range of failure locations, timings, and durations, providing broad coverage of the failure-state space without explicit enumeration. We acknowledge that this approach does not exhaustively cover all node-level failure combinations, and that nodes with poor connectivity or high demand may warrant targeted coverage. Future extensions could employ importance sampling or stratified failure sampling to ensure that such critical nodes are adequately represented in the scenario set.} To simulate these failures, a stochastic scenario generation framework is employed, in which outage realizations are sampled according to the scenario probabilities derived from the fault tree analysis and applied to the representative days obtained through clustering. 

\subsection{Final Scenario Construction}

The complete scenario set is constructed by combining representative days with failure scenarios. For each representative day $d_{\text{rep}}$ with weight $w_{\text{rep}}$, three scenarios are generated with adjusted weights:\vspace{-5mm}
\begin{align}
w_{\text{scenario},1} &= P_{\text{normal}} \times w_{\text{rep}}, \\
w_{\text{scenario},2} &= P_{\text{single}} \times w_{\text{rep}}, \\
w_{\text{scenario},3} &= P_{\text{combined}} \times w_{\text{rep}}.
\end{align}
The total number of scenarios is given by:\vspace{-2mm}
\begin{equation}
N_{\text{scenarios}} = 3 \times \sum_{s=1}^{4} K_s^*,
\label{eq:total_scenarios}
\end{equation}
where $K_s^*$ is the optimal number of clusters selected for season $s$. Summing over all clusters and scenario types, the total weight across all scenarios can be expressed as:

\begin{equation}
\begin{aligned}
\sum_{i=1}^{N_{\text{scenarios}}} w_i &= \sum_{d} w_{\text{rep},d} \times \left( P_{\text{normal}} + P_{\text{single}} + P_{\text{combined}} \right)\\ 
&= \sum_{d} w_{\text{rep},d} \times 1 = 1.
\label{eq:weight_normalization}
\end{aligned}
\end{equation}

\hl{For each representative day $d$, the operating states corresponding to normal operation, single failure, and combined failures are mutually exclusive and collectively exhaustive, such that their probabilities sum to unity. Moreover, the weights $w_{\text{rep},d}$ denote the relative contribution of each representative day to the annual horizon and are normalized to sum to one. Therefore, aggregating over all representative days yields a total scenario weight of unity, ensuring probabilistic consistency of the scenario set. This property is verified computationally to maintain the probabilistic consistency of the scenario set.}

 Table~\ref{Table 2} presents the seasonal distribution of clusters, total number of representative days, and the resulting number of final scenarios (three per representative day). The minimum number of clusters per season, $K$, is a key design parameter that governs the fidelity of the scenario representation. To determine an appropriate value, we performed a convergence analysis by solving the full optimization problem for increasing values of $K$, starting from $K = 2$ and incrementing through $K = 3, 4, 5, 6, 7$. As reported in Table~\ref{Tab.3}, the objective function value stabilizes at $K \geq 7$, with negligible change observed between 72 and 90 scenarios ($K_{\min} = 5$ versus $K_{\min} = 7$). This convergence behaviour indicates that seven clusters per season provide sufficient operational diversity to capture the essential variability in load and PV patterns without introducing redundant scenarios. Accordingly, the final configuration adopts $K = 7$ as the minimum number of clusters per season, yielding a total of 30 representative days and 90 scenarios (three failure modes per representative day), which are used for all subsequent analysis.
\begin{table*}[!th]
\centering
\scriptsize
\caption{Summary of clustering configurations and final scenario set sizes.}
\label{Table 2}
\renewcommand{\arraystretch}{1.3}
\setlength{\tabcolsep}{4pt}
\begin{tabular}{|
>{\centering\arraybackslash}p{2.5cm}|
>{\centering\arraybackslash}p{4.2cm}|
>{\centering\arraybackslash}p{2.2cm}|
>{\centering\arraybackslash}p{3cm}|
}
\hline
\textbf{Min K per Season} & \textbf{Seasonal K (Summer, Autumn, Winter, Spring)} & \textbf{Representative days ($\sum K_s$)} & \textbf{Final Scenarios (3 $\times$ Representative days)} \\
\hline
2 & 2-2-2-2 & 8  & \textbf{24} \\
3 & 3-3-3-4 & 13 & \textbf{39} \\
4 & 4-4-4-4 & 16 & \textbf{48} \\
5 & 6-7-7-6 & 26 & \textbf{78} \\
6 & 7-7-8-7 & 29 & \textbf{87} \\
7 & 7-7-8-8 & 30 & \textbf{90} \\
\hline
\end{tabular}
\end{table*}

\begin{figure*}[!t] 
    \centering 
    \includegraphics[width=0.9\linewidth]{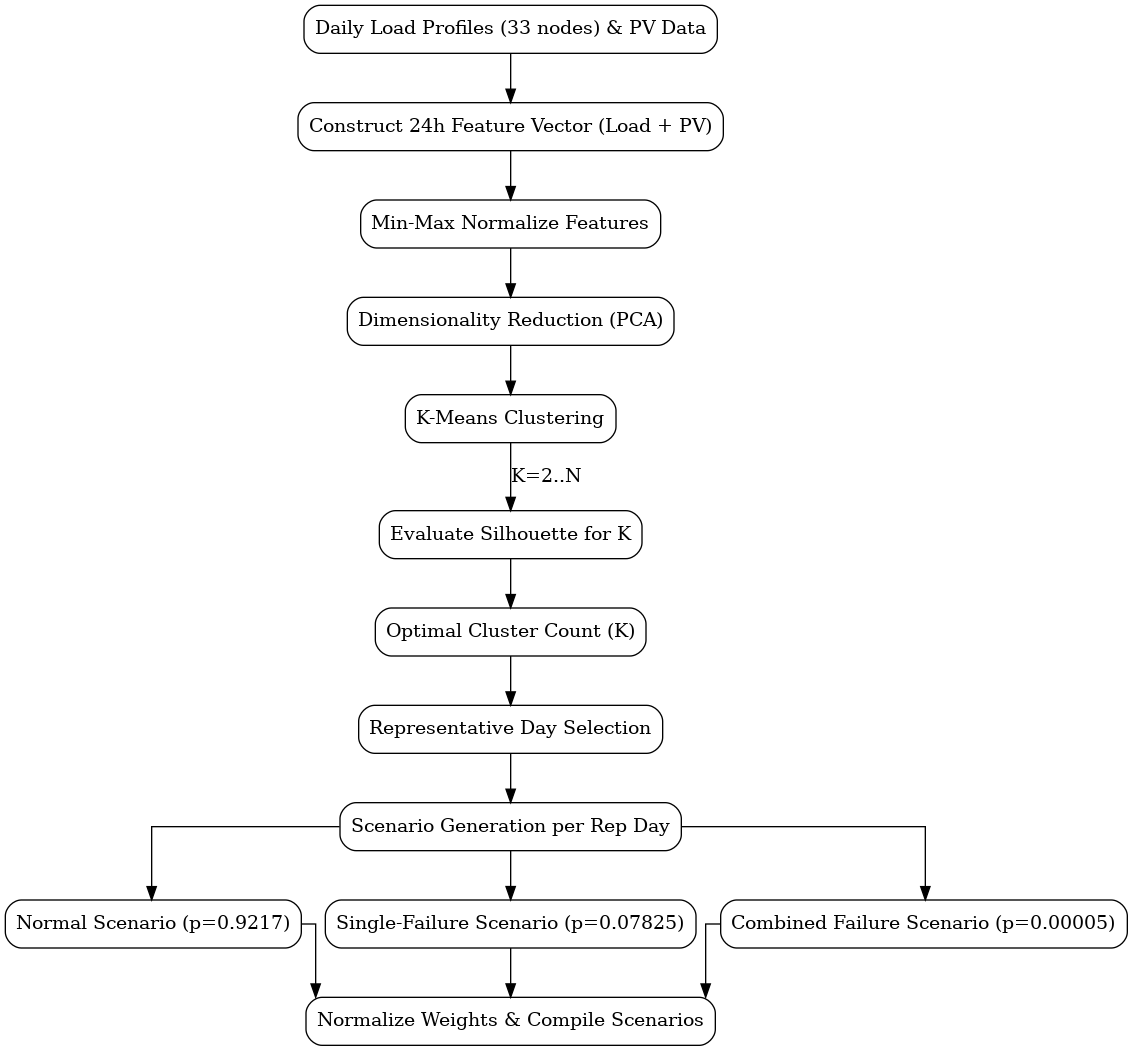} 
    \caption{Scenario generation methodology.} 
    \label{Fig.3}  
\end{figure*}

Fig.~\ref{Fig.3} presents the complete scenario generation pipeline adopted in this work. The process begins with the construction of daily feature vectors that concatenate the 24-hour load profiles of all nodes with the corresponding PV generation profile. These feature vectors are then stratified by season, normalized, and reduced in dimensionality via PCA to enable efficient clustering. K-means clustering is applied within each season to identify representative days, with cluster weights reflecting annual significance. Finally, each representative day is combined with the three mutually exclusive failure scenarios (normal operation, single-component failure, and combined failure) derived from the fault tree analysis, producing weighted operational scenarios that collectively capture renewable intermittency, demand variability, and equipment failures.

The scenario generation methodology described in this section produces a compact yet representative set of operational scenarios that systematically accounts for the principal sources of uncertainty in off-grid microgrid operation. The resulting scenario set serves as the foundation for the mathematical optimization framework presented in the following section, which co-optimizes investment and operational decisions subject to reliability and network constraints.

\section{Mathematical Formulation}

In this section, the optimization framework formulation for reliability-oriented microgrid is introduced. Despite being solved as a single integrated problem, the decision variables exhibit a temporal hierarchy. Planning decisions, such as investing in PV and battery capacities, must be made in advance of uncertainty realizations. These "here-and-now" decisions are fixed across all scenarios and define the feasible space within which the system can operate. In contrast, operational decisions, such as power dispatch, energy-not-served (ENS), and storage management, are made after uncertainties unfold. These "wait-and-see" decisions are scenario dependent and allow the system to respond adaptively to different conditions. This structure captures the trade-off between investment robustness and operational flexibility under uncertainty.

\subsection{Objective Function}

The objective function integrates three components over the planning horizon, whereas the first two components capture planning stage specific considerations, while the third one addresses operational aspects. The first component (1) covers the initial capital expenditure for PV units, and batteries, independent of operational scenarios. The second component (2) reflects the operation \& maintenance (O\&M) costs over the planning horizon, discounted to present value using rate $r$, and directly proportional to installed capacities. This capacity-based O\&M modeling is standard practice in long-term energy planning studies, where O\&M expenditures are predominantly driven by the installed asset base (e.g., routine inspections, panel cleaning, and battery conditioning) rather than by the volume of energy dispatched. \hl{Walker et al.~\cite{walker2020model} confirmed this relationship in a detailed PV O\&M cost model showing that maintenance activities scale with installed capacity rather than generation output. Similarly, Quashie et al.~\cite{quashie2018optimal} adopted the same capacity-proportional O\&M formulation in their microgrid planning framework, further validating its appropriateness for long-term investment optimization.} Hence, modeling O\&M in this manner is appropriate at the planning stage because it captures the dominant fixed-maintenance component of lifecycle costs while maintaining linearity in the optimization, and avoids coupling O\&M with scenario-dependent dispatch decisions that would substantially increase computational complexity. Finally, the third component (3) incorporates operational reliability through ENS, which is weighted by scenario probabilities $\pi_s$. \hl{Supply interruptions are penalized using the Value of Lost Load (VOLL), set to 24.86 AUD/kWh as adopted from the AEMC Value of Customer Reliability study~\cite{AER2024VCR}. It is important to note that, in the presence of the binding hard reliability constraint, this VOLL penalty serves primarily as a regularizer that governs how the optimizer distributes the allowable ENS across nodes and time periods, provides a smooth cost gradient that aids solver convergence, and ensures meaningful ENS penalization should the reliability threshold be adjusted for alternative planning contexts.} The integrated objective function is given below: 
\vspace{-3mm}

\begin{align*}
\min \quad & \underbrace{\sum_{i \in \mathcal{N}} (c^\text{pv} \times \text{cap}_i^\text{pv} + c^\text{b} \times \text{cap}_i^\text{b})}_\text{Investment Cost} + \\
& \underbrace{\sum_{n=1}^H \frac{1}{(1+r)^n} \times \left[\sum_{i \in \mathcal{N}} (c^\text{O\&M}_\text{pv} \times \text{cap}_i^\text{pv} + c^\text{O\&M}_\text{b} \times \text{cap}_i^\text{b})\right]}_\text{Present Value of O\&M Costs} + \\
& \underbrace{\sum_{n=1}^H \frac{1}{(1+r)^n} \times \left(\sum_{s \in \mathcal{S}} \pi_s \left[\sum_{t \in \mathcal{T}} \sum_{i \in \mathcal{N}} \text{VOLL}_i \times \text{ENS}_{i,t,s} \times \Delta t\right]\right)}_\text{Present Value of EENS cost}
\end{align*}

\subsection{Constraints}

The following set of constraints forms the foundation of our two-stage stochastic optimization model, ensuring the physical feasibility of PV and battery installations while meeting operational limits under uncertainty. Each constraint addresses specific aspects of the 100\% renewable system: planning constraints limit node-specific capacities based on space and technical bounds, while operational constraints govern power dispatch, storage dynamics, and network flows across scenarios capturing intermittency and failures. Although many of these constraints represent standard operational and physical limits for microgrids, this work introduces a probabilistic ENS constraint that explicitly enforces utility-grade reliability standards (e.g.. 99.998\% demand fulfillment) in a 100\%-renewable context.

\subsubsection{Planning Stage Constraints}
The constraints of the planning stage limit investment decisions as follows: 
\begin{equation}
\text{cap}_i^\text{pv} + (p^\text{max} \times \text{cap}_i^\text{b}) \leq S^\text{base}_i, \quad \forall i \in \mathcal{N}.
\label{eq:node_capacity}
\end{equation}

In Eq.~\eqref{eq:node_capacity}, the total installed capacities of PV and battery at each node are subject to the capacity of the node. Therefore, they are restricted by the maximum allowed capacity for that node.  The decision variable \(\text{cap}_i^{\text{b}}\) denotes the optimized \emph{energy} capacity of the battery at node \(i\) (in kWh). The parameter \(p^{\max}\) (set to \(1~\mathrm{kW}/\mathrm{kWh}\) in this study) specifies the fixed C\,-\,rate that links the battery’s admissible \emph{power} rating to its energy capacity. This coupling is a physically meaningful design constraint: it ensures that the instantaneous power the battery can deliver (or absorb) scales with its installed energy. Accordingly, the product of \(p^{\max}\, and \mathrm{cap}^{\mathrm{b}}_i\) represents the battery’s power capacity (in kW) and enters the node capacity constraint to correctly account for the battery’s contribution to the total allowable power at bus \(i\).

Eq.~\eqref{eq:pv_cap_limit} incorporates, for each node $i$, the installed PV capacity ${cap}_i^\text{pv}$ must not exceed the maximum allowed PV capacity ${cap}_\text{pv}^\text{max}$ as determined. 

\begin{equation}
\text{cap}_i^\text{pv} \leq \text{cap}_\text{pv}^\text{max}, \quad \forall i \in \mathcal{N}.
\label{eq:pv_cap_limit}
\end{equation}

Similarly, eq.~\eqref{eq:batt_cap_limit} ensures that the battery capacity at each node does not exceed the maximum allowable battery capacity, considering technical and safety limitations.

\begin{equation}
\text{cap}_i^\text{b} \leq \text{cap}_\text{b}^\text{max}, \quad \forall i \in \mathcal{N}.
\label{eq:batt_cap_limit}
\end{equation}

\subsubsection{Operational Constraints}

The operational constraints govern resource dispatch across all scenarios. The operational constraint below limits the PV power output based on the available solar resource and equipment status, considering potential failures.

\begin{equation}
0 \leq p_{i,t,s} \leq \text{pv}_{t,s}^\text{g} \times \text{cap}_i^\text{pv} \times (1-x_{pv,i,s}^\text{fail}), \quad \forall i \in \mathcal{N}, \forall t \in \mathcal{T}, \forall s \in \mathcal{S}.
\label{eq:pv_output}
\end{equation}

Eq.~\eqref{eq:batt_power} governs both charging and discharging limits through symmetrical power limitations, which are proportionally scaled according to the installed capacity while accounting for potential failures.

\begin{equation}
-p^\text{max} \times \text{cap}_i^\text{b} \times (1-x_{b,i,s}^\text{fail}) \leq p_{i,t,s}^\text{b} \leq p^\text{max} \times \text{cap}_i^\text{b} \times (1-x_{b,i,s}^\text{fail}), ~ \forall i,t,s.
\label{eq:batt_power}
\end{equation}

The state of charge dynamics represented in Eq.~\eqref{eq:soc_dynamics} is one of the most complex temporal relationships in the microgrid optimization model.  For time periods after the initial period (t > 1), this constraint captures the evolution of the battery's stored energy over time. The first term ${soc}_{i,t-1,s}^\text{b} \times (1-x^{fail}_{b,i,s})$ represents the energy carried over from the previous time period, accounting for potential battery failures through the binary parameter $x^{fail}_{b,i,s}$. The second term handles both charging and discharging operations: $\max(0,p^b_{i,t,s})$ isolates positive values representing discharge, which is then divided by the discharge efficiency $\eta^{dis}$ to account for energy losses during discharge, while $\min(0,p^b_{i,t,s})$ isolates negative values representing charging, which is multiplied by the charging efficiency $\eta^{ch}$ to account for losses during charging. These terms are multiplied by the time step duration $\Delta t$ to convert power to energy, and by $(1-x^{fail}_{b,i,s})$ to ensure no charging or discharging occurs if the battery has failed. 

For $t > 1$:
\begin{equation}
    \begin{aligned}
        &{soc}_{i,t,s}^\text{b}\\
        &=\left[{soc}_{i,t-1,s}^\text{b}-\left(\frac{1}{\eta^{dis}}\,\max(0,p^b_{i,t,s})+\eta^{ch}\,\min(0,p^b_{i,t,s})\right)\Delta t\right]\left(1-x^{fail}_{b,i,s}\right), \\
        &\forall i \in \mathcal{N}, \forall t \in \mathcal{T}\setminus\{1\}, \forall s \in \mathcal{S}.
        \label{eq:soc_dynamics}
    \end{aligned}
\end{equation}

For the initial time period ($t = 1$), the recursive relationship in Eq.~\eqref{eq:soc_dynamics} requires a boundary condition, since the preceding state of charge lies outside the modeled operational horizon. In practice, each battery begins the representative day with a state of charge inherited from the previous day's operation. However,  as this model enforces a cyclic SoC constraint that equates the terminal and initial states of charge, the initialization value is self-consistent: the battery begins each day at the same level to which it returns by the end of that day. This initialization is specified as follows:


For $t = 1$:
\begin{equation}
\text{soc}_{i,1,s}^\text{b} = \text{soc}_{b,i}^\text{init}, \quad \forall i \in \mathcal{N}, \forall s \in \mathcal{S}.
\label{eq:soc_init}
\end{equation}

Hence, Eq.~\eqref{eq:soc_init} sets the initial state of charge for battery operation at each node. This initialization is crucial as it provides the necessary boundary condition for the recursive relationship that governs the battery's state of charge throughout the operational horizon. In this study, the initial SoC fraction $soc^{\mathrm{init}}_{b,i}$ is set to a neutral mid-level value (50\% of energy capacity) to avoid biasing the dispatch toward either charging or discharging-dominated schedules.  Importantly, the influence of this initial value on the optimization outcome is mitigated by two mechanisms. First, the cyclic state-of-charge constraint (Eq.~\eqref{eq:soc_cyclic}) requires that the terminal SoC at $t = T$ equals the initial SoC at $t = 1$, thereby eliminating end-of-horizon artifacts and ensuring that the chosen initial level is sustained in steady-state operation. Second, the optimization is free to adjust the battery dispatch trajectory throughout the day to accommodate any initial condition; consequently, the objective function value is insensitive to moderate variations in $\text{soc}_{b,i}^{\text{init}}$, provided the cyclic constraint is enforced. As a result, the 50\% initialization serves as a practical and neutral starting point that does not bias the investment or operational decisions.

In addition, Eq.~\eqref{eq:soc_bounds} maintains battery state of charge within acceptable limits by preventing deep discharge and avoiding overcharging, ensuring safe operation in all scenarios.
\begin{equation}
\text{soc}^\text{min} \times \text{cap}_i^\text{b} \leq \text{soc}_{i,t,s}^\text{b} \leq \text{soc}^\text{max} \times \text{cap}_i^\text{b}, \quad \forall i,t,s.
\label{eq:soc_bounds}
\end{equation}

The nodal power balance at each bus is enforced by:
\begin{equation}
p_{i,t,s} + p_{i,t,s}^\text{b} - \sum_{j:(i,j) \in \mathcal{E}} P_{ij,t,s} + \sum_{j:(j,i) \in \mathcal{E}} P_{ji,t,s} + \text{ENS}_{i,t,s} = d_{i,t,s}, \quad \forall i,t,s.
\label{eq:power_balance}
\end{equation}

Eq.~\eqref{eq:power_balance} balances the generation, storage, and power flows at each node, while allowing load shedding through the ENS variable when local and imported supply is not sufficient. Moreover, the following constraint enforces line capacity limits by:

\begin{equation}
-P_{ij}^\text{max} \leq P_{ij,t,s} \leq P_{ij}^\text{max}, \quad \forall (i,j) \in \mathcal{E}, t, s.
\label{eq:line_capacity}
\end{equation}

To ensure acceptable power quality, the nodal voltages are bounded within standard operational limits:
\vspace{-3mm}
\begin{equation}
V_{\min}^2 \;\leq\; V_{i,t,s} \;\leq\; V_{\max}^2,
\quad \forall i \in \mathcal{N},\; t \in \mathcal{T},\; s \in \mathcal{S}.
\label{eq:voltage_bounds}
\end{equation}

Moreover, this work adopts a linearized DistFlow (LinDistFlow) approximation to represent the voltage behavior along the radial distribution feeder. The LinDistFlow model provides an appropriate level of fidelity for distribution level planning problems because it captures the dominant physical constraint governing voltage profiles in radial networks, namely the resistive voltage drop due to active power flow, while maintaining a linear formulation that enables tractable large scale stochastic optimization with provable global optimality. Although LinDistFlow can be optimistic under heavy loading conditions by neglecting quadratic loss terms, this approximation has been shown to introduce negligible error in distribution systems operating within standard voltage limits. \hl{Low~\cite{low2014convex} provided a rigorous theoretical foundation by deriving the conditions under which the linear DistFlow relaxation is exact for radial networks, establishing the basis for its application in distribution level planning. Building on this theoretical grounding, Scalfati et al.~\cite{scalfati2017optimal} subsequently corroborated its practical utility within a MILP based DER sizing framework for smart microgrids, demonstrating that the linearization error remained within practically acceptable bounds.} The accuracy of this approximation is further improved in microgrids with distributed generation that reduces long-distance power transfers. Furthermore, given that modern inverter-based DERs provide rapid local reactive power compensation, this work assumes reactive power is autonomously regulated at each node and decouples it from the voltage optimization problem. Under this framework, the voltage drop across each line $(i,j)\in\mathcal{E}$ is expressed as:
\vspace{-3mm}
\begin{equation}
V_{j,t,s}
=
V_{i,t,s}
-
\frac{2 r_{ij}}{V_{\text{base}}^2 \cdot 1000}
P_{ij,t,s},
\quad \forall (i,j)\in\mathcal{E},\; t\in\mathcal{T},\; s\in\mathcal{S}.
\label{eq:lindistflow_forward}
\end{equation}

To account for bidirectional power flow enabled by distributed energy resources and storage systems, the reverse-direction voltage relationship is also enforced:
\begin{equation}
V_{i,t,s}
=
V_{j,t,s}
-
\frac{2 r_{ij}}{V_{\text{base}}^2 \cdot 1000}
P_{ji,t,s},
\quad \forall (i,j)\in\mathcal{E},\; t\in\mathcal{T},\; s\in\mathcal{S}.
\label{eq:lindistflow_reverse}
\end{equation}

Since the LinDistFlow approximation characterizes voltage differences rather than absolute magnitudes, a reference point must be established. The first node is therefore selected as the slack bus for the islanded microgrid:
\begin{equation}
V_{1,t,s} = 1.0, 
\quad \forall t\in\mathcal{T},\; s\in\mathcal{S}.
\label{eq:slack_bus}
\end{equation}

Together, \eqref{eq:voltage_bounds}--\eqref{eq:slack_bus} provide a computationally efficient yet sufficiently accurate representation of voltage behavior for distribution-level microgrid planning, preserving linearity and ensuring global optimality of the mixed-integer formulation.

Furthermore, the following constraint ensures the cyclic operation of batteries by matching the end and start states of charge.
\begin{equation}
\text{soc}_{i,T,s}^\text{b} = \text{soc}_{i,1,s}^\text{b}, \quad \forall i \in \mathcal{N}, \forall s \in \mathcal{S}
\label{eq:soc_cyclic}
\end{equation}

\hl{The above constraint enforces cyclic battery operation by requiring that each battery ends the representative day at the same state of charge (SoC) at which it begins. This requirement arises from the use of independent representative days, which do not preserve chronological continuity across periods. In the absence of inter-day linkage, stored energy has no value beyond the end of each representative day, and the optimization would otherwise favor complete discharge at the terminal time step. Enforcing cyclic SoC therefore prevents end-of-horizon artifacts and ensures that each representative day is energetically self-consistent and can be interpreted as a repeatable daily operating pattern within the planning horizon.

From a modeling perspective, the SoC is tracked over $T+1$ states for $T$ operational time steps, with the inter-temporal dynamics (Eq. 24) linking hourly charging and discharging decisions to subsequent energy states. In addition, physically infeasible behavior is avoided by preventing simultaneous charging and discharging at any node through the adopted formulation, ensuring realistic battery operation across all scenarios.}

To ensure the physical feasibility of the solution, all physical quantities are constrained to be non-negative:
\begin{equation}
p_{i,t,s} \geq 0, \quad \text{soc}_{i,t,s}^\text{b} \geq 0, \quad \text{ENS}_{i,t,s} \geq 0, \quad \forall i,t,s.
\label{eq:nonneg}
\end{equation}

Finally, the probabilistic system-wide reliability constraint that explicitly quantifies and bounds the  EENS across all operational scenarios is formulated as: 
\begin{equation}
\sum_{s \in \mathcal{S}} \pi_s \left[\sum_{t \in \mathcal{T}} \sum_{i \in \mathcal{N}} \text{ENS}_{i,t,s} \times \Delta t\right] \leq \epsilon_{rel} \times \sum_{s \in \mathcal{S}} \pi_s \left[\sum_{t \in \mathcal{T}} \sum_{i \in \mathcal{N}} d_{i,t,s} \times \Delta t\right].
\label{eq:reliability}
\end{equation}

Eq~\eqref{eq:reliability} enforces reliability by computing the EENS as a probability-weighted aggregation across all scenarios $s \in \mathcal{S}$, all time periods $t \in \mathcal{T}$, and all nodes $i \in \mathcal{N}$. The left-hand side therefore captures both the spatial dimension (system-wide ENS summed over every node and hour) and the probabilistic dimension (weighted by scenario probabilities $\pi_s$ to reflect the likelihood of each operational condition, including normal operation and component failure events). The resulting EENS is then required to remain below a fraction $\epsilon_{rel}$ of the similarly weighted expected annual demand on the right-hand side. \hl{In this study, we set $\epsilon_{rel} = 0.00002$ (0.002\%) to adopt the reliability threshold established by the Australian Energy Market Commission (AEMC)~\cite{AEMC2024} as a demanding design benchmark. While the AEMC standard was originally defined for grid-connected systems with interconnection and reserve sharing, no equivalent regulatory reliability standard currently exists for isolated off-grid microgrids in Australia. Hence, we adopt this threshold not as a claim of regulatory compliance, but as a conservative design target that ensures the microgrid delivers service quality comparable to conventional grid supply. This choice is motivated by the principle that remote communities transitioning from grid-connected service, or aspiring to grid-equivalent reliability, should not be expected to accept a lower standard simply because their supply is off-grid. Moreover, because this threshold represents the most stringent commonly referenced benchmark in the Australian context, a framework that satisfies it can trivially accommodate less demanding reliability requirements. Importantly, this threshold is a configurable parameter, allowing planners to adjust it to match local regulatory requirements or community-specific reliability expectations.}

In summary, the mathematical formulation presented in this section constitutes a two-stage stochastic mixed-integer linear program that co-optimizes investment and operational decisions for a 100\%-renewable microgrid. The planning-stage constraints (Eqs.~\eqref{eq:node_capacity}--\eqref{eq:batt_cap_limit}) bound the installed capacities of PV and battery units at each node, while the operational-stage constraints (Eqs.~\eqref{eq:pv_output}--\eqref{eq:nonneg}) govern resource dispatch, storage dynamics, and network power flows across all scenarios. The probabilistic reliability constraint (Eq.~\eqref{eq:reliability}) ties these two stages together by requiring that the scenario-weighted EENS remains within the prescribed regulatory threshold. The resulting formulation is solved as a single integrated optimization problem, and its performance is evaluated in the following section through numerical simulations on the IEEE 33-bus test system.

\section{Results \& Discussion}

To validate the proposed framework, the IEEE 33-bus test system is employed to simulate the operation of a 100\% renewable energy-based microgrid. Applying the stochastic scenario generation methodology described in Section~3.6, the scenario probabilities for this case study are obtained as $P_{\text{normal}} = 0.9217$, $P_{\text{single}} = 0.07825$, and $P_{\text{combined}} = 0.00005$. The microgrid exhibits a peak load demand of 663.33 kW, with a daily average energy consumption of 4316.30 kWh/day. The optimization problem is implemented in Python 3.13 and solved using the CPLEX solver on a computer with an Intel Core(TM) i7-12800H @2.40 GHz processor, 32 GB RAM, and a 64-bit operating system.

\subsection{Convergence Analysis and Scenario Configuration}

To determine the appropriate number of representative days, a convergence analysis was conducted across multiple clustering configurations. \hl{Table~\ref{Tab.3} reports the objective function value and average solve time for each clustering configuration. Each configuration was solved five times with different solver random seeds; the objective value was identical across all runs for each configuration, confirming solution robustness. The average solve times are non-monotonic across configurations. This is a well-documented property of mixed-integer programming, where solver performance depends on the tightness of the LP relaxation and the structure of the branch-and-bound tree rather than on problem size alone. Smaller scenario sets can produce a looser LP relaxation that requires deeper branching to close the optimality gap, whereas additional scenarios can tighten the feasible region and improve solver bounds, reducing computation despite the larger problem size. All configurations solve within approximately three minutes on average, confirming that the 90-scenario formulation remains computationally tractable for a planning-stage problem solved once during the design phase.

The convergence trajectory reveals that premature stabilization can occur at low values of K. The objective remained nearly unchanged between K = 2 (8.16M AUD, 24 scenarios) and K = 3 (8.17M AUD, 39 scenarios), which might initially suggest convergence. However, increasing to K = 4 shifted the objective to 8.33M AUD, and a further increase at K = 5 raised it to 8.97M AUD, demonstrating that the earlier apparent stability was an artifact of insufficient operational diversity in the scenario set. To ensure genuine convergence, the analysis was continued through K = 6 and K = 7, both of which returned an identical objective of 8.97M AUD. The observation of three consecutive configurations with the same objective value provides robust assurance that the 90-scenario design has achieved true convergence and is not subject to the premature stabilization observed at lower values of K. This convergence criterion, combined with the silhouette-constrained clustering analysis described in Section 3.4, jointly justify K = 7 as the selected minimum cluster threshold.}

\begin{table}[htbp]
\centering
\scriptsize
\caption{\hl{Convergence validation across scenario sets.}}
\label{Tab.3}
\renewcommand{\arraystretch}{1.2}
{
\begin{tabular}{cccc}
\toprule
\textbf{K (min/season)} & \textbf{Total Scenarios} & \textbf{Solve Time (s)} & \textbf{Objective (M\$)} \\
\midrule
K=2 & 24 & 177 & 8.16 \\
K=3 & 39 & 29  & 8.17 \\
K=4 & 48 & 45  & 8.33 \\
K=5 & 78 & 146  & 8.97 \\
K=6 & 87 & 197 & 8.97 \\
K=7 & 90 & 166  & 8.97 \\
\bottomrule
\end{tabular}
}
\end{table}

\subsection{Seasonal Load and PV Profiles}

\begin{figure}[!t] 
    \centering 
    \includegraphics[width=1\linewidth]{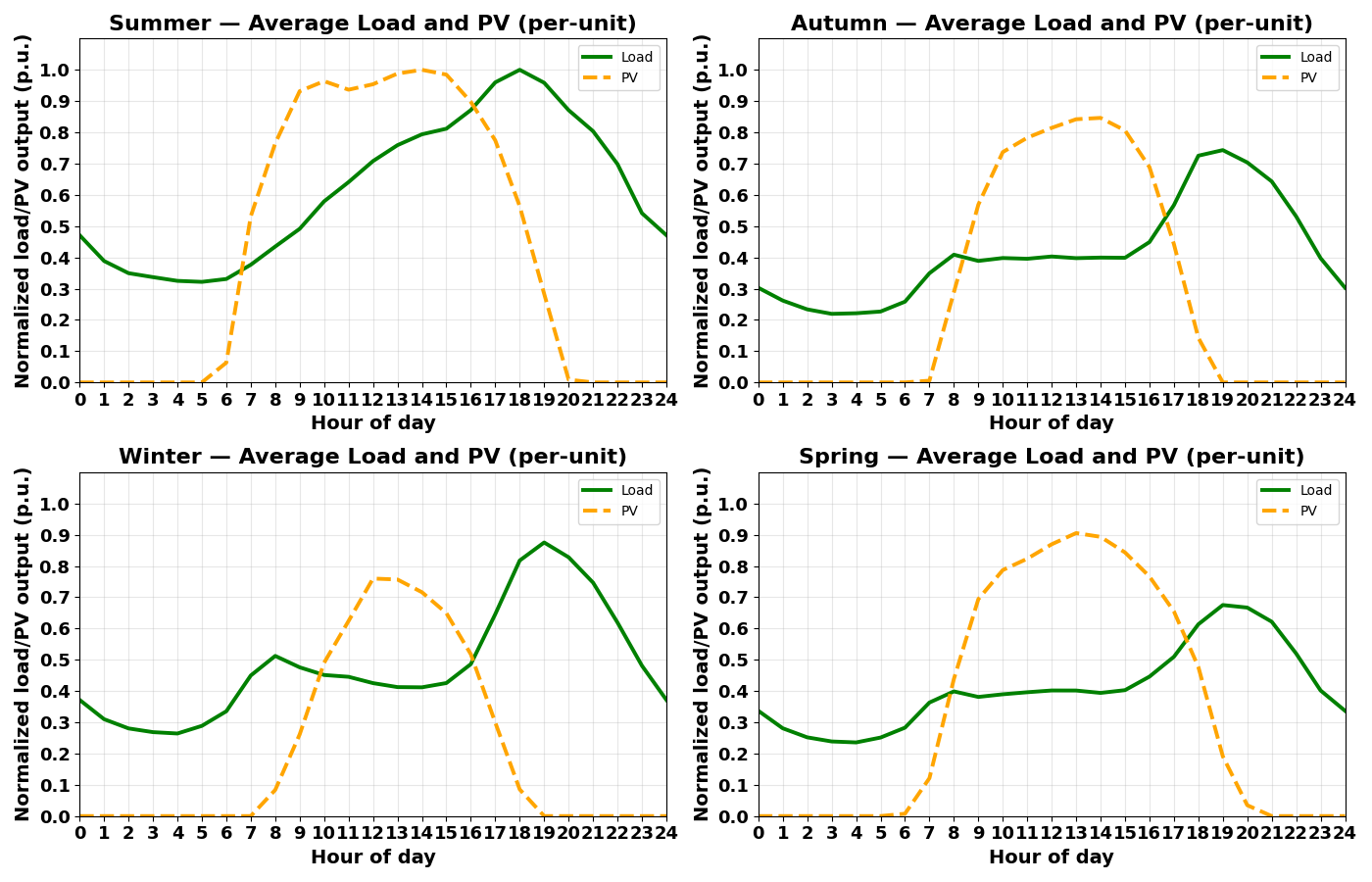} 
    \caption{Seasonal average load and PV profile in normalized per-unit form. \parbox{0.95\linewidth}{\centering \textit{Note: The PV profile represents normalized output per unit of installed capacity, while the load series is normalized to the annual peak demand to enable direct comparison of temporal patterns on a common dimensionless scale.}}}
    \label{Fig.4}  
\end{figure}

The hourly average system load profile, together with the normalized PV generation profile (per unit) for spring, summer, autumn, and winter, is illustrated in Fig.~\ref{Fig.4}. It highlights strong seasonal dependencies in the balance between renewable supply-demand. In spring and summer, abundant solar availability conveniently coincides with daytime load requirements, meaning that PV generation is often high during peak load hours. By contrast, autumn and winter present more challenging conditions: solar production is significantly reduced during the shorter, cloudier days, whereas evening loads are higher. This mismatch, particularly in winter, characterized by lower PV input and elevated demand after sunset, emphasizes the importance of sufficient storage and network support to maintain reliability. From a planning perspective, this seasonal asymmetry implies that the winter season is the binding constraint for both storage sizing and reliability: the optimizer must provision sufficient battery capacity to bridge the prolonged overnight gap in winter, even though this capacity remains partially underutilized during the more favorable spring and summer months. Consequently, winter adequacy requirements constitute a major contributing factor to the cost of achieving utility-grade reliability in a 100\% renewable microgrid.

\subsection{Optimized Resource Allocation}

\begin{figure}[!t] 
    \centering 
    \includegraphics[width=0.9\linewidth]{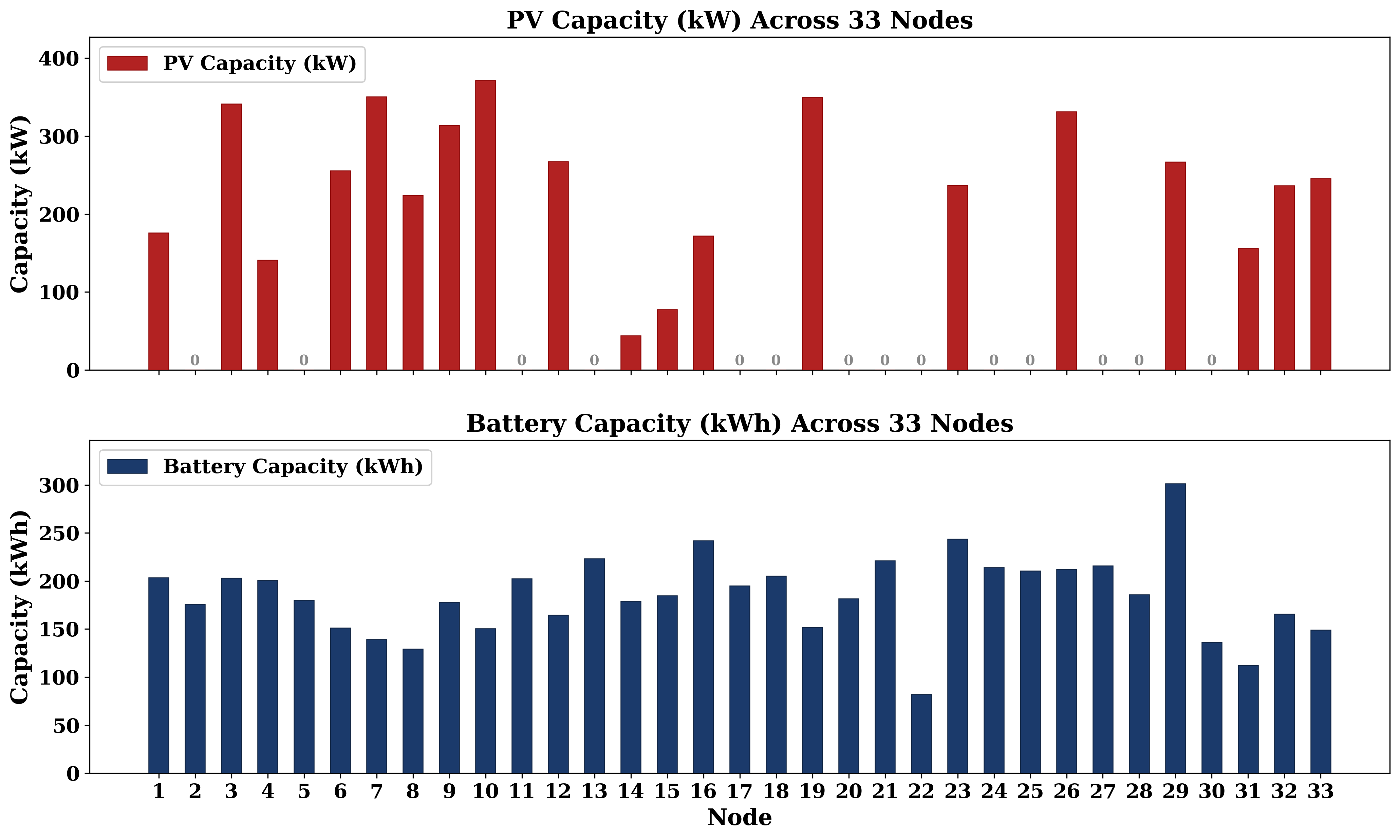}
    \caption{\hl{PV and battery capacity across the system.}} 
    \label{Fig.5}  
\end{figure}

The optimization results yield a distributed placement of PV generation and battery storage across the 33-bus network, as presented in Fig.~\ref{Fig.5}. A distinct allocation pattern emerges, in which battery storage is deployed at every node with capacities ranging from 82~kWh at Node 22 to 301~kWh at Node 29, yielding a total system-wide storage capacity of 6,091~kWh. In contrast, PV generation is installed at only 19 of the 33 nodes, with 14 nodes receiving no local PV despite the optimizer's freedom to place generation at any bus. The total installed PV capacity reaches 4,558~kW, concentrated at nodes where it contributes most effectively to network-level power sharing. 

This selective allocation reflects a deliberate design logic, whereby storage serves as a universal reliability buffer at every bus to ensure that each node retains local capacity to bridge overnight gaps and absorb short-term supply disruptions, while PV generation is strategically concentrated at nodes where it can serve both local demand and export surplus energy to the wider network. Nodes without local PV rely entirely on energy imports from PV-rich nodes combined with their local battery storage for time-shifted supply. The bias toward universal battery deployment stems from a fundamental operational requirement of a 100\% renewable system, since solar generation alone cannot meet demand at all hours, and the optimization therefore prioritizes distributed storage to ensure that every node can sustain supply during nighttime and periods of low solar generation or PV outages. This resulting allocation is strategically asymmetric rather than uniform or load-proportional, underscoring the value of topology-aware planning in identifying cost-effective resource configurations that maintain reliability across the entire network.

\begin{figure}[!t] 
    \centering 
    \includegraphics[width=1\linewidth]{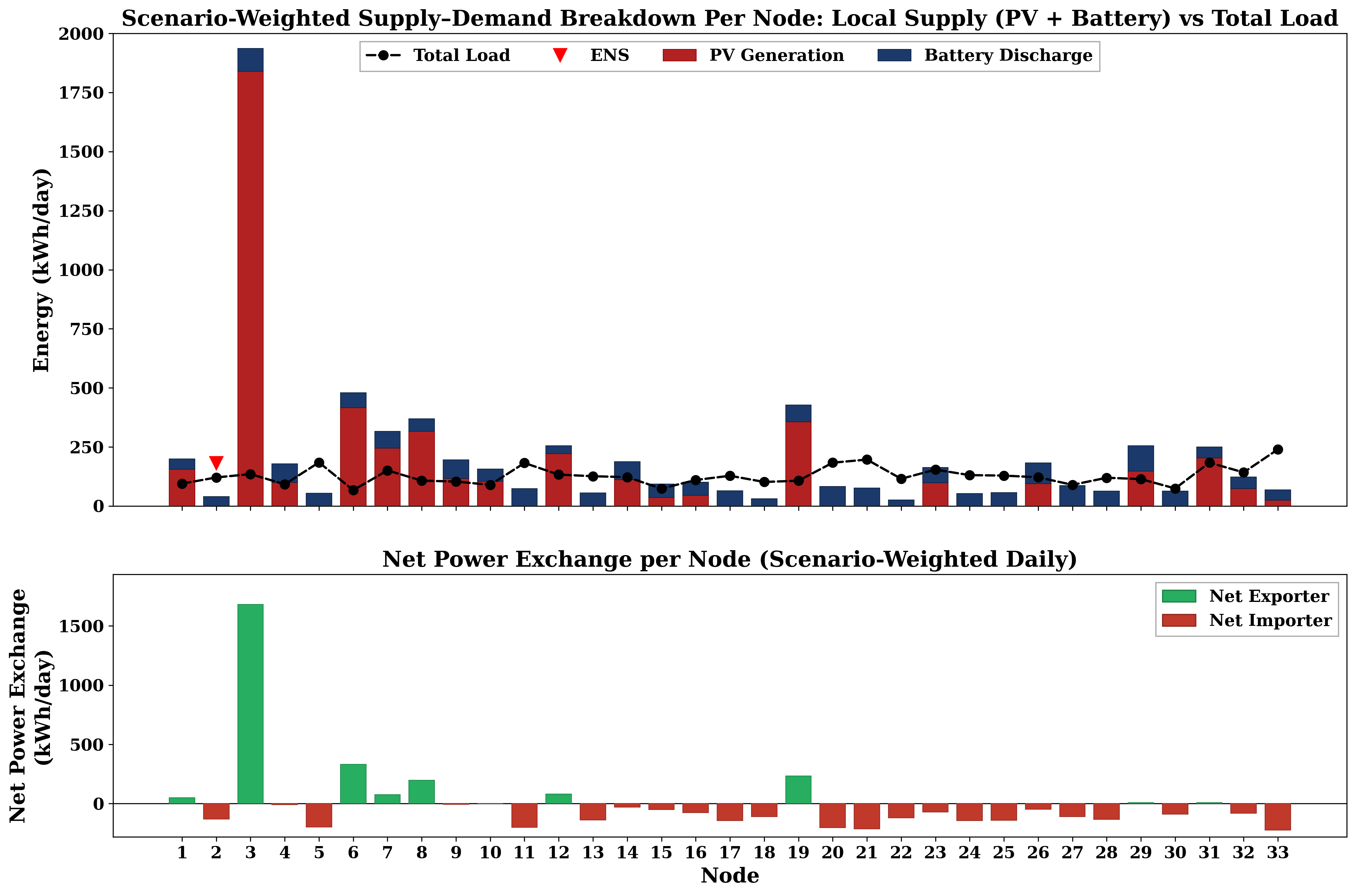} 
    \caption{\centering \hl{Scenario-weighted nodal energy balance across the IEEE 33-bus microgrid, showing PV generation, battery discharge, load demand, net power exchange, and EENS per bus.}  {\parbox{0.95\linewidth}{\centering \textit{Note: Negative values in the net power exchange panel indicate net importers that rely on power inflows from neighboring nodes.}}}}
    \label{Fig.6}  
\end{figure}

A scenario-weighted nodal energy balance analysis, presented in Fig.~\ref{Fig.6} reveals the spatial distribution of energy generation, consumption, and inter-nodal power sharing across the microgrid network. The model identifies a clear functional differentiation among nodes, in which 10 nodes operate as net exporters that supply surplus energy to the network, while the remaining 23 nodes are net importers that depend on power inflows from neighboring nodes to fully meet their demand.

Node 3 emerges as the dominant generation hub, dispatching approximately 1,840~kWh/day of PV energy against a local load of only 135~kWh/day and thereby producing a net export of 1,682~kWh/day to the rest of the network, by far the largest contribution of any single node. Additional generation anchors include Node 6 with 333~kWh/day net export, Node 19 with 233~kWh/day, and Node 8 with 197~kWh/day. These PV-rich nodes transmit surplus energy through the distribution lines to sustain nodes that lack sufficient local generation. At these exporting nodes, the stacked local supply consisting of PV generation and battery discharge substantially exceeds the local load, as shown in the top panel, with the excess flowing outward through the network.

On the importing side, Node 33 is the largest net importer at 225~kWh/day, followed by Node 21, Node 20, Node 11, and Node 5. Notably, among these five nodes, Nodes 21, 20, 11, and 5 are entirely without local PV and rely on imported energy combined with their local battery storage for time-shifted supply. Node 33 presents a particularly instructive case. Despite having 246~kW of installed PV, it remains the system's largest net importer because its local load of 240~kWh/day, the highest in the network, exceeds its scenario-weighted PV dispatch. The PV capacity at this node is sized to ensure local adequacy during critical low-solar or failure scenarios rather than to meet average daily demand, and during typical operation the deficit is covered by imports from generation-rich nodes. This reliability-driven sizing logic is similarly observed at other PV-equipped nodes that are nevertheless net importers, including Node 26, Node 23, and Node 32.

Across the entire simulation horizon, total daily exports and imports balance at approximately 2,674~kWh, confirming system-wide energy conservation. The absence of unmet load at 32 of the 33 nodes, despite 23 nodes being net importers, demonstrates that the distributed resource allocation and network connectivity are sufficient to ensure demand fulfillment throughout the system. The sole exception is Node 2, which experiences a marginal annual ENS of approximately 30.83~kWh. This dynamic exchange of energy underscores the critical role of topology-aware planning, as the optimizer exploits the network's capacity to transport energy between nodes, concentrating PV generation where it is most productive and relying on power sharing to serve nodes where local generation alone would be insufficient.

\subsection{System-Level Operational Profiles}

\begin{figure}[!t] 
    \centering 
    \includegraphics[width=0.95\linewidth]{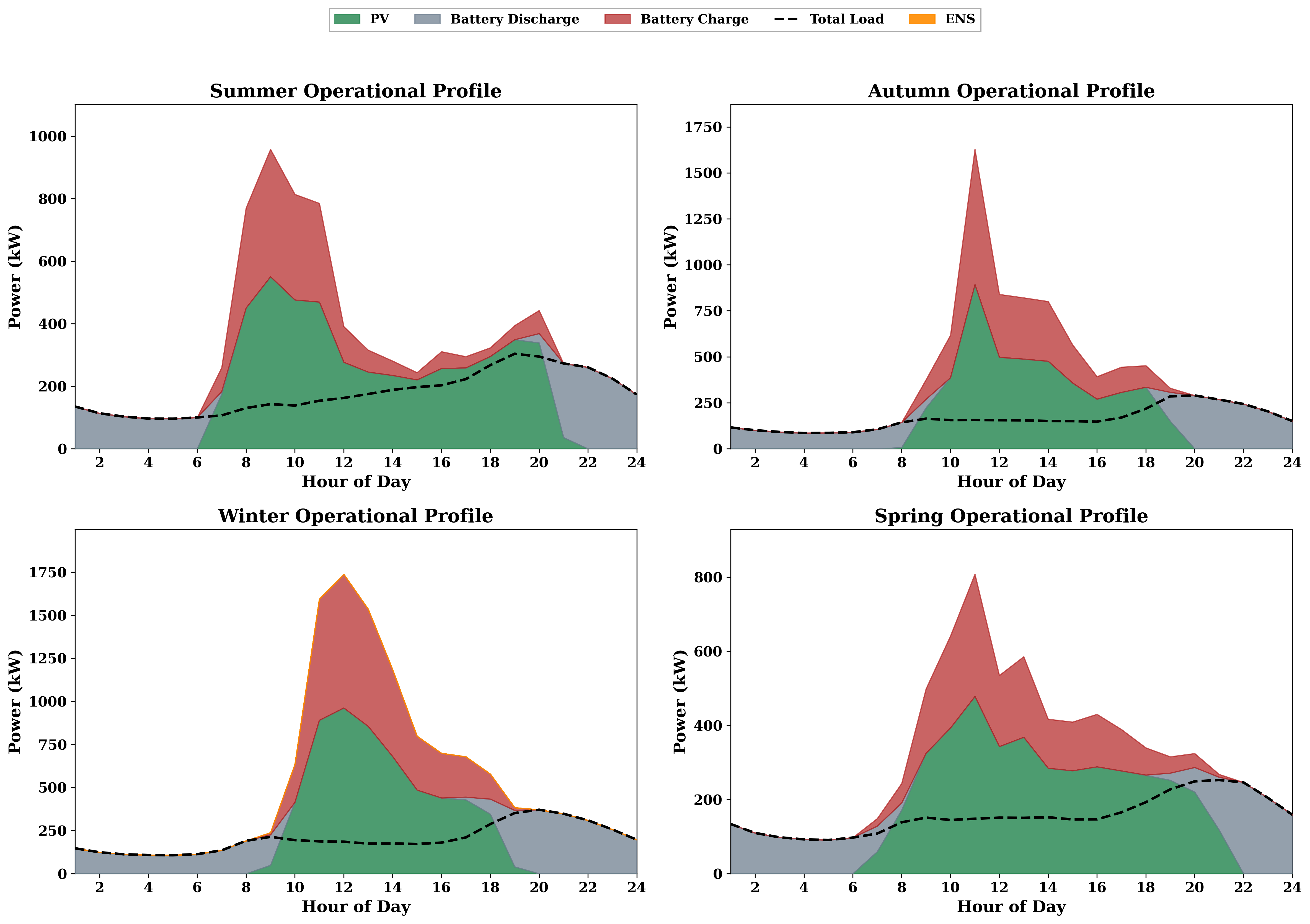} 
    \caption{\centering Scenario-weighted hourly operational profiles for each season, showing system-level PV generation, battery charging and discharging, load demand, and EENS}
    \label{Fig.7}  
\end{figure}

Operation across all scenarios reveals a consistent daily cycling pattern for generation, storage, and load that remains structurally stable across seasons, as shown in Fig.~\ref{Fig.7}. During overnight and early morning hours (approximately hours 22--7), when solar output is zero, the microgrid relies predominantly on battery discharging to supply the base load, with average system-wide discharge rates ranging from approximately 127 kW in autumn to 161 kW in winter. As dawn breaks, PV generation progressively takes over from battery discharge, with the transition occurring around hour 7 in summer and spring and slightly later in autumn and winter. By mid-morning, PV output covers a substantial portion of the demand in all seasons, and any production exceeding the instantaneous load is harnessed to charge the battery fleet. This behavior is evident from the pronounced midday charging activity, with peak charging power ranging from 330--410~kW in summer and spring to 740--780~kW in autumn and winter. The higher charging rates in autumn and winter reflect an operational strategy to maximize the capture of available solar energy during shorter sunshine periods, ensuring that batteries recover sufficient state of charge to bridge the prolonged overnight gap.

Following the decline in afternoon solar production, the storage units transition back into discharge mode to supply the evening and late-night loads. This diurnal pattern of charging when the sun is up and discharging when it is down is consistently observed in all seasons, with only the magnitude of charging/discharging and the timing of transitions shifting in accordance with daylight hours. This consistency of the diurnal dispatch pattern across seasons is a noteworthy finding, as it indicates that the optimized microgrid follows a structurally stable operating pattern despite significant seasonal variation in resource availability. Although this charge–discharge behavior aligns with general expectations for solar-storage systems, the optimization results confirm that it remains robust under the modeled reliability and adequacy constraints. From a practical perspective, this suggests that a simple rule-based dispatch approach, in which storage is charged during solar production hours and discharged during non-solar hours, can approximate the optimized solution while reducing operational complexity for remote community operators without access to advanced optimization tools.

\hl{Importantly, the dispatch profiles exhibit smooth and physically consistent transitions through the final hour of the day across all four seasons. At hour 24, battery discharge matches the load precisely, with no simultaneous charging and discharging activity at the system level. This behavior confirms the correct operation of the cyclic state-of-charge constraint (Eq. 33), which requires each battery to return to its initial state of charge by the end of the representative day. This is a standard technique in representative-day energy system optimization, ensuring that daily dispatch patterns are self-consistent and repeatable across the planning horizon.}

Crucially, the microgrid maintains optimized daily supply--demand balance while satisfying the stringent reliability criteria imposed in the design. The EENS remains effectively zero across nearly all seasons and hours, confirming that the probabilistic reliability constraint, which bounds the allowable fraction of unmet demand, is met by the optimized design. The EENS curve remains near zero throughout the year-long simulation, demonstrating adherence to the annual reliability target. 

\subsection{Node-Level Failure Analysis}

\begin{figure}[!t] 
    \centering 
    \includegraphics[width=1\linewidth]{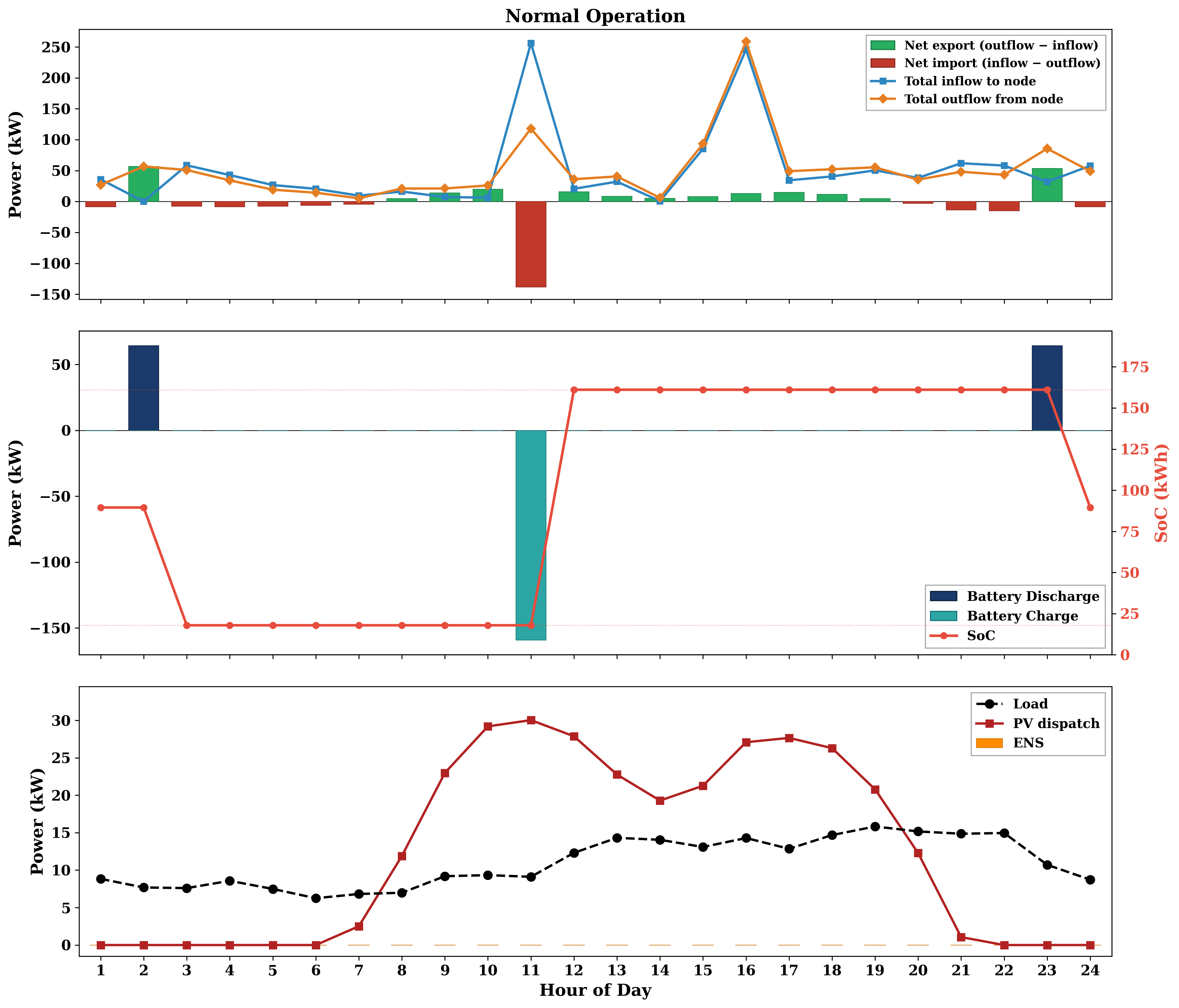} 
    \caption{\centering \hl{Hourly operational profile of Node~14 under normal operation (no component failures): Net power exchange with the network (Top), Battery charge/discharge along with state of charge (Middle), PV dispatch with load demand, and ENS (Bottom).}}
    \label{Fig:Normal}  
\end{figure}

\begin{figure}[!t] 
    \centering 
    \includegraphics[width=1\linewidth]{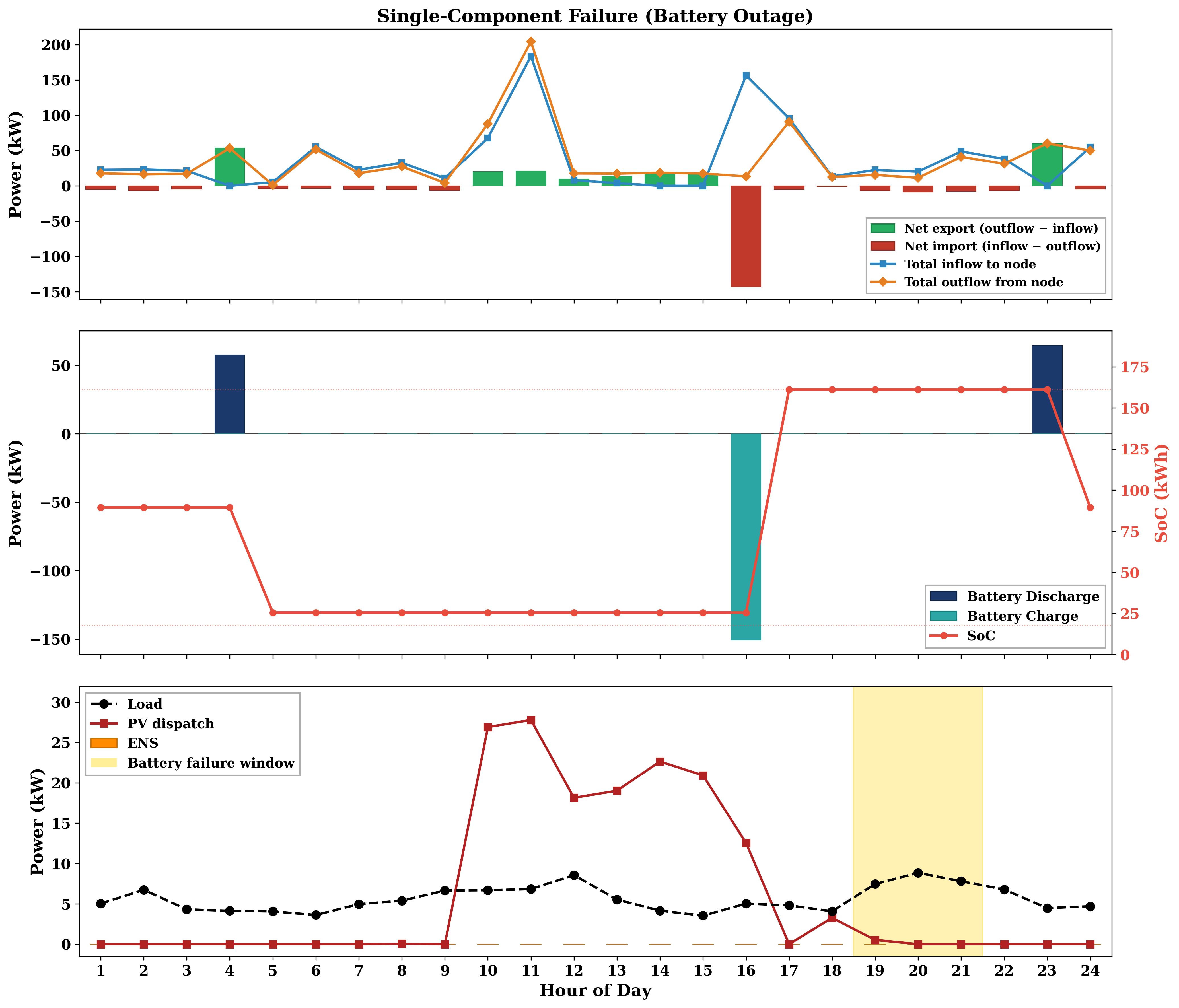} 
    \caption{\centering \hl{Hourly operational profile of Node~14 under a single-component failure scenario: Net power exchange with the network (Top), Battery charge/discharge along with state of charge (Middle), PV dispatch with load demand (including failure window), and ENS (Bottom).}} 
    \label{Fig:Single Outage}  
\end{figure}

\begin{figure}[!t] 
    \centering 
    \includegraphics[width=1\linewidth]{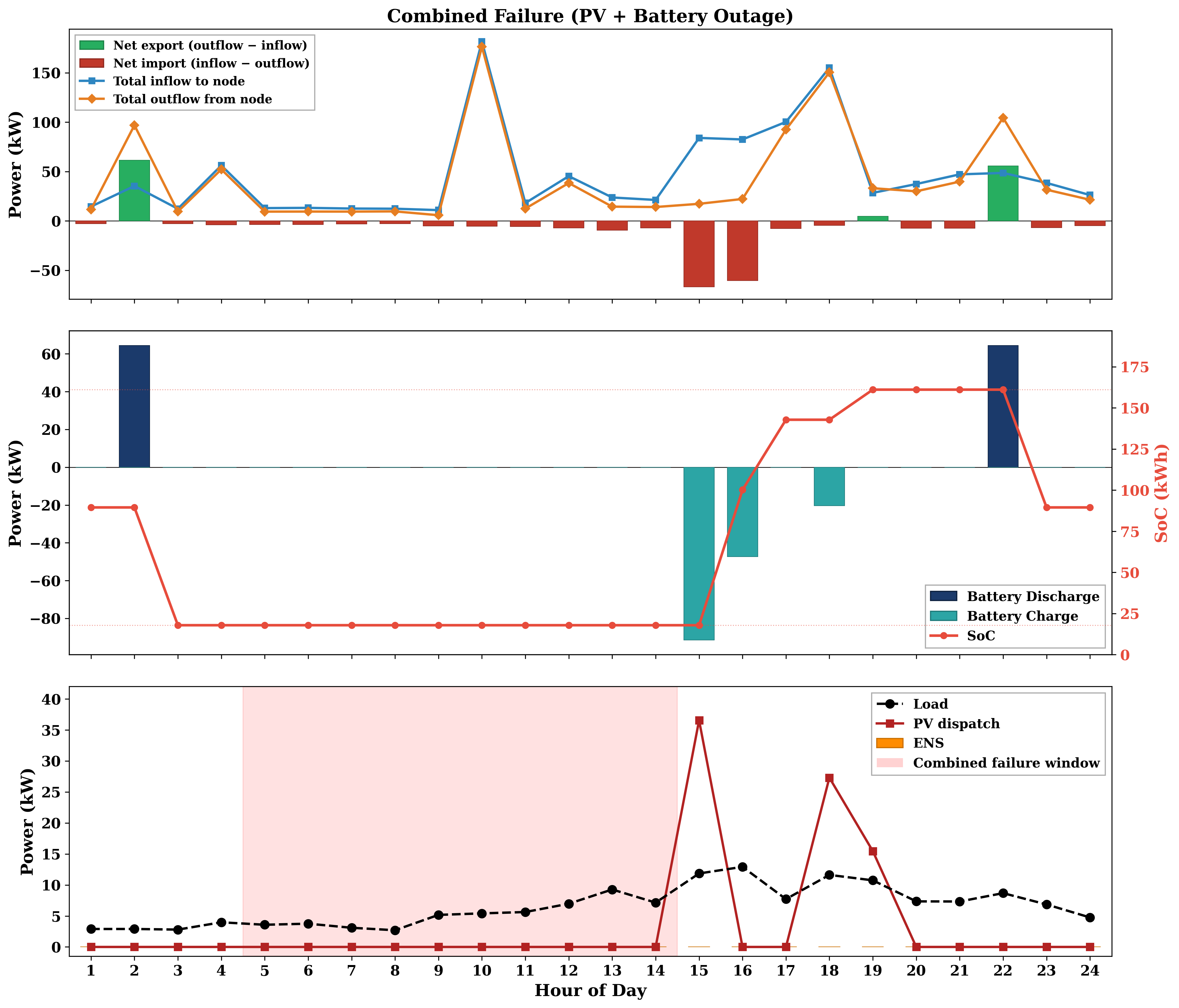} 
    \caption{\centering \hl{Hourly operational profile of Node~14 under a combined-component failure scenario: Net power exchange with the network (Top), Battery charge/discharge along with state of charge (Middle), PV dispatch with load demand (including failure window), and ENS (Bottom).}} 
    \label{Fig:Combined Outage}  
\end{figure}

The optimal microgrid configuration obtained from the proposed optimization framework is evaluated under several failure scenarios for key components. Fig.~\ref{Fig:Normal}, Fig.~\ref{Fig:Single Outage}, and Fig.~\ref{Fig:Combined Outage}, respectively focuses on Node 14 as a representative case, depicting its power operation under (i) normal operation, (ii) battery outage, and (iii) simultaneous PV-and-battery outage. 

Under normal operating conditions, Node 14 functions as a mildly import-dominated node. During the night, the node imports power from the network to meet its base load, and its battery remains at a low state of charge until mid-morning. At approximately hour 11, a large charging event of 159~kW rapidly replenishes the battery. Although the node's local PV output of 30~kW exceeds its load of 9~kW at this hour, the local surplus alone is insufficient to sustain charging at this rate. The remaining charging power is therefore drawn from the network, exploiting the system-wide PV surplus available during peak solar hours. This coordinated behavior, in which energy is imported for local storage when the broader network has excess generation, demonstrates how the optimizer leverages temporal complementarity between distributed generation and storage across the network. From midday through the late afternoon (hours 12--19), the node's local PV dispatch exceeds its load, briefly turning the node into a net exporter. As PV output declines in the late afternoon, the node reverts to net importing. The battery discharges in two distinct events, at hour 2 and hour 23, to support both local demand and the wider network during overnight hours. The state of charge begins and ends the day at 89.5~kWh, confirming compliance with the cyclic SoC constraint, and the ENS remains zero throughout the day. Overall, normal operation demonstrates how a node coordinates network imports and strategic battery cycling with local generation to maintain uninterrupted supply.

In the second scenario, Node 14 experiences a battery outage over a three-hour window spanning hours 19 to 21, during which battery operation is completely suppressed. Throughout this period, both charging and discharging are forced to zero, and the state of charge remains frozen at its pre-failure value of 161~kWh. Despite the loss of local storage during evening hours when PV output is negligible, Node 14 does not experience any supply disruption or load curtailment. Instead, the node imports its entire demand from neighboring nodes through the distribution network, as reflected by the net import values recorded during the failure window. Prior to the outage, the battery charges at hour 16 with 151~kW to build up stored energy. After the outage clears at hour 22, the battery resumes operation with a large discharge event of 64~kW at hour 23, exporting stored energy back to the network while returning to the required end-of-day state of charge.
This scenario demonstrates that the loss of local storage at a single node has negligible operational impact, since upstream PV and battery capacity elsewhere in the microgrid seamlessly compensate for the outage. The result carries an important design implication, in that the proposed distributed planning approach inherently embeds single--component resilience into the microgrid without requiring explicit contingency constraints. Because the optimizer sizes resources to satisfy reliability targets across all failure scenarios simultaneously, the resulting design naturally tolerates single--component outages through network--level redundancy rather than through local over--provisioning at each bus.

The third scenario examines a combined PV and battery outage at Node 14 lasting 10 hours, from hour 5 to hour 14, representing a worst-case localized failure in which both generation and storage are simultaneously unavailable. During this window, PV dispatch is forced to zero and the battery SoC is frozen at its pre-failure value of 17.9~kWh, with no charging or discharging permitted. As a result, the node becomes purely demand-driven, and in each failure hour the entire load is supplied through net imports from neighboring nodes, with the inflow closely tracking the local demand. Despite this simultaneous loss of both local generation and storage across the core solar hours when PV would normally contribute, the ENS remains zero in every hour, confirming that the rest of the microgrid retains sufficient capacity and network connectivity to fully support Node 14 throughout the outage.

Once the failure clears at hour 15, the optimization adjusts the operating pattern of Node 14's assets to recover. A rapid battery charging sequence occurs at hours 15 and 16, with charging powers of 91~kW and 47~kW respectively, followed by an additional charge of 20~kW at hour 18, rebuilding the SoC from 17.9~kWh to 161~kWh. The battery subsequently discharges 64~kW at hour 22 to return to the required end-of-day state of charge while simultaneously exporting stored energy to support the wider network. Similarly, PV dispatch resumes immediately once the failure clears. At hour 15, PV generation rises sharply to 36.6~kW, its highest output of the day, as the node capitalizes on the remaining afternoon solar window to simultaneously serve local load, charge the depleted battery, and export surplus energy to the network. A secondary PV contribution of 27.3~kW at hour 18 further supports the battery recovery before solar output ceases for the evening.

This combined failure scenario underscores the resilience of the distributed microgrid design to correlated outages. Even when a node loses all local support for an extended period spanning the primary solar generation window, the redundancy of the system and the network-level power sharing prevent any interruption in supply. These results highlight the critical role of topology-aware distributed resource planning, as the spatial diversity of distributed energy resources across the network acts as a built-in redundancy layer that ensures the temporary loss of a single node's resources does not propagate into broader system instability. \hl{To further demonstrate the robustness of the distributed design, an additional failure analysis is presented in Appendix~A for Node 13, a battery-only node with no local PV. This analysis confirms that even nodes without any local generation maintain uninterrupted supply during extended equipment outages, relying solely on network-level power sharing.}

\subsection{Reliability Performance and ENS Analysis}

\begin{figure*}[!t] 
    \centering 
    \includegraphics[width=0.8\linewidth]{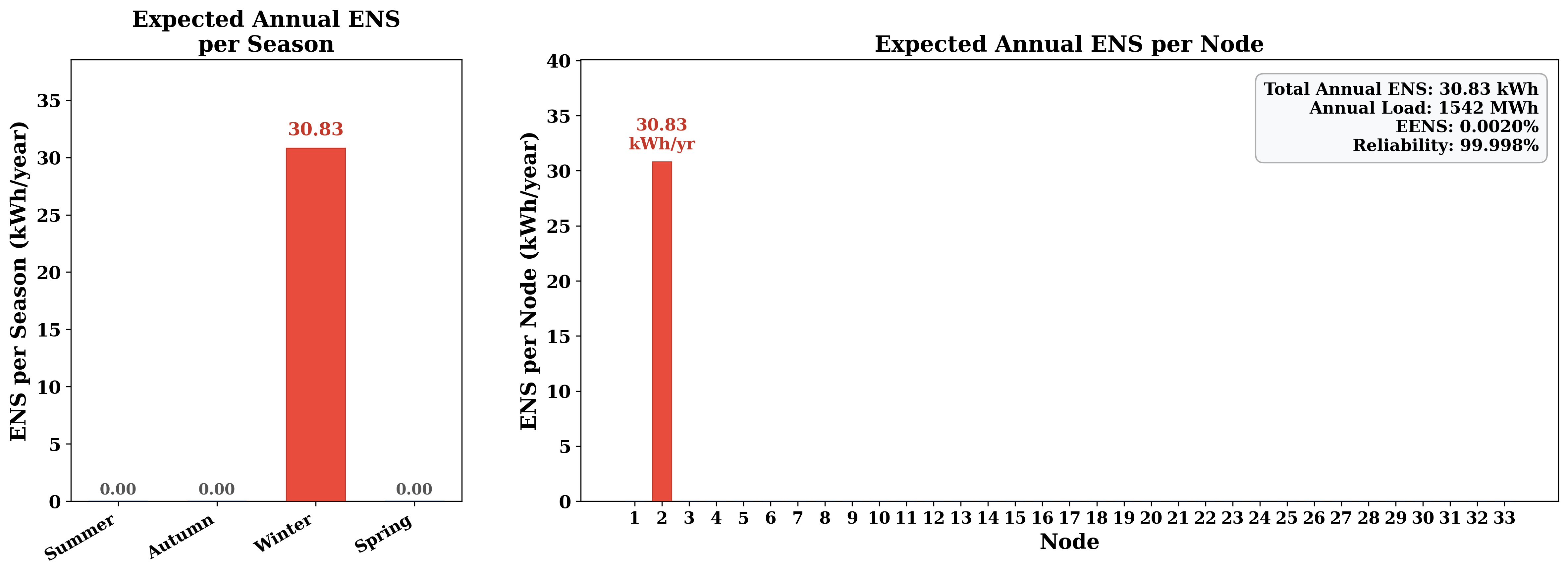} 
    \caption{\hl{ENS per season and per node.}} 
    \label{Fig.12}  
\end{figure*}

\hl{The simulation results confirm that the probabilistic reliability constraint embedded in the optimization effectively limits EENS across all operational conditions. Across all 90 scenarios and 24 hourly time periods, EENS is zero for every node and season except winter, where a small but non-zero shortfall occurs under the most demanding combination of low solar availability and elevated heating loads, as shown in Fig.~\ref{Fig.12}. The total annual shortfall is approximately 30.83 kWh, concentrated entirely at Node 2. This node lacks local PV and depends on network imports combined with its local battery (176 kWh) for supply. During critical winter periods, reduced solar generation across the network limits the energy available for import, and the stored energy at Node 2 is insufficient to fully bridge the extended overnight gap, leading to brief supply deficits.}

\hl{While this shortfall remains well within the design threshold (0.002\% of annual demand), it represents the binding reliability constraint and identifies winter adequacy as the critical bottleneck for purely solar-battery microgrids. In remote community contexts where even brief supply interruptions may have disproportionate consequences, this finding carries practical significance: it pinpoints precisely where, and when (e.g., winter) supplementary resources might provide the greatest marginal reliability improvement. The mathematical formulation is inherently technology-agnostic, and such resources can be incorporated by introducing corresponding decision variables and constraints without modifying the optimization structure.

From a system perspective, the microgrid achieves 99.998\% supply reliability, demonstrating that 100\%-renewable off-grid microgrids can deliver service quality aligned with the AEMC benchmark adopted in this study. The concentration of the residual shortfall in a single season and at a single node, rather than distributed broadly across the network, further validates the effectiveness of the topology-aware distributed planning approach in maintaining supply security under diverse operating conditions.}

\subsection{Comparison with Baseline Designs }
\hl{To quantify the value of the proposed framework, a centralized single--node baseline is evaluated. In the centralized model, all loads and distributed energy resources are aggregated into a single node, eliminating network topology, line flow constraints, and node-specific investment decisions. This representation corresponds to the single--node modeling approach adopted by the majority of existing microgrid planning studies, as documented in Table~\ref{tab:comparison}. Same scenario generation methodology and cost parameters have been applied to ensure a fair comparison.}

\begin{table*}[htbp]
\centering
\caption{\hl{Comparison of distributed (proposed) and centralized microgrid designs.}}
\label{tab:comparison1}
 {
\begin{tabular}{lcc}
\hline
\textbf{Metric} & \textbf{Distributed (Proposed)} & \textbf{Centralized} \\
\hline
Total PV capacity (kW) & 4,558 & 20,051 \\
Total battery capacity (kWh) & 6,091 & 5,685 \\
Total NPV (M AUD) & 8.97 & 19.52 \\
Annual ENS (kWh/year) & 30.83 & 4,663.62 \\
EENS (\%) & 0.002 & 0.302 \\
Supply reliability (\%) & 99.998 & 99.698 \\
Reliability target met & EENS $\leq$ 0.002\% $\checkmark$ & EENS $\leq$ 0.002\% $\times$ \\
\hline
\end{tabular}}
\end{table*}

\hl{Table~\ref{tab:comparison1} reveals a striking and fundamental limitation of the centralized modeling approach. The centralized design is infeasible at the 99.998\% reliability target, the same target that the distributed design satisfies at a total system cost of 8.97~M~AUD. In fact, the centralized model remains infeasible even when the reliability requirement is progressively relaxed to 99.9\% and 99.8\%, achieving feasibility only at 99.7\% (EENS $\leq$ 0.3\%), a reliability level that permits 150 times more unserved energy than that achieved by the proposed distributed design. Even at this substantially relaxed target, the centralized design costs 19.52~M~AUD, more than double the cost of the proposed design, which achieves far superior reliability.

The capacity allocation reveals the source of this cost disparity. The centralized design installs 20,051~kW of PV, 4.4 times the 4,558~kW deployed in the distributed design, yet still fails to match its reliability performance. This substantial PV oversizing is a direct consequence of the single-node architecture, since without network topology to distribute resources across spatially separated nodes, the centralized model must provision enough generation at the single aggregated point to cover demand under all operating conditions, including worst-case failures. The battery capacities, by contrast, are comparable between the two designs, at 5,685~kWh for the centralized model and 6,091~kWh for the distributed model, indicating that the storage requirement is driven primarily by the diurnal energy--shifting need, which is similar regardless of network representation.

The reliability gap is equally striking. The centralized design produces 4,664~kWh/year of unserved energy, 151 times greater than the 30.83~kWh/year achieved by the distributed design, despite costing more than twice as much. This disparity arises from a fundamental structural limitation of the centralized model, in that a combined failure event, in which both PV and battery at the aggregated node fail simultaneously, removes all local generation and storage from the entire system in a single contingency. With no spatial redundancy or power-sharing capability from unaffected nodes, the centralized model has no recourse mechanism to serve load during such events, making stringent reliability targets unachievable regardless of the installed capacity. By contrast, the distributed design spreads resources across 33 nodes, ensuring that a combined failure at any individual node affects only a fraction of the total system capacity. The remaining healthy nodes continue to operate and share power through the network, maintaining supply continuity at the affected node, as demonstrated in the failure analyses of Figs.~8 to 10.

These results provide definitive evidence that the topology-aware distributed approach is not merely a modeling refinement but a structural enabler of high reliability. The centralized single-node assumption fundamentally misrepresents the resilience characteristics of distributed microgrids by eliminating the spatial diversity that enables network-level power sharing during component outages. The distributed design simultaneously achieves 150 times lower ENS, meets a reliability target that the centralized model cannot satisfy at any cost, and does so at less than half the investment, a combination of outcomes that underscores the critical importance of incorporating network topology into the microgrid planning process.}

\subsection{Adaptation to Real-World Microgrids}

\hl{While the case study in this work employs the IEEE 33-bus test system as the network topology, the practical value of any optimization-based planning framework ultimately depends on its ability to transition from a standardized test environment to a real-world microgrid setting. To apply the proposed framework on an actual remote Australian microgrid requires only a set of clearly defined input-layer adaptations, which are outlined below:

(1) \textbf{Network topology and line data:} The IEEE 33-bus feeder configuration should be replaced with the actual network topology and line impedance data obtained from the relevant distribution network service provider (DNSP). The only structural requirement is that the network retains a radial configuration, which is the standard topology for remote Australian distribution feeders.

(2) \textbf{Demand profile:} The synthetic load profiles assigned to each bus should be substituted with site-specific metered demand data obtained from smart meters or supervisory control and data acquisition (SCADA) systems, thereby capturing the actual temporal consumption behavior of the community.

(3) \textbf{Renewable generation profile:} The renewable energy output profiles should be updated using site-specific resource data sourced from the Bureau of Meteorology or relevant platform.

(4) \textbf{Reliability threshold:} The reliability constraint threshold should be adjusted to align with the applicable regulatory reliability standard or the service-level expectation defined by the community.

Finally, it is worth emphasizing that these substitutions are confined entirely to the input data layer. The underlying optimization model, stochastic scenario structure, and solution algorithm, remains unchanged regardless of the specific site to which it is applied.}

\section{Conclusion}
This research addresses the critical challenge of designing cost-effective 100\% renewable microgrids that maintain utility-grade reliability standards. The proposed scenario-based joint optimization framework represents a paradigm shift from traditional cost-centric approaches by embedding explicit reliability constraints directly into the planning process. Through comprehensive integration of long-term investment decisions with short-term operational strategies under uncertainty, the methodology ensures robust microgrid designs capable of handling renewable intermittency, equipment failures, and demand fluctuations.

The validation results demonstrate that distributed network-aware resource allocation, combined with probabilistic scenario modeling, enables the achievement of 99.998\% supply reliability using exclusively renewable resources. \hl{While this threshold is adopted from the AEMC benchmark as a demanding design target to demonstrate that 100\%-renewable off-grid microgrids can achieve reliability levels comparable to grid-connected supply}, the framework is designed to be jurisdiction-agnostic: the reliability threshold $\epsilon_{rel}$ in Eq.~\eqref{eq:reliability} is a configurable parameter that can be adjusted to match alternative regulatory benchmarks, such as the North American Electric Reliability Corporation (NERC) Loss of Load Expectation (LOLE) standards, the European Network of Transmission System Operators for Electricity (ENTSO-E) adequacy targets, or region-specific electrification requirements in developing economies where reliability expectations may differ. Because the reliability constraint is expressed as a fractional bound on EENS relative to total demand, it accommodates any threshold without requiring structural changes in the optimization formulation. The optimized microgrid design maintains electricity supply continuity comparable to conventional grid service, with minimal energy shortfalls occurring only under the most demanding winter conditions. This performance validates the technical and economic feasibility of diesel-free 100\%-renewable microgrids for remote and rural electrification applications across diverse regulatory contexts.

Key contributions include the development of a holistic planning methodology that treats reliability as a primary design criterion rather than a secondary consideration, innovative scenario generation techniques that capture real-world uncertainties, and network topology integration that optimizes distributed resource placement. The framework provides energy planners and policymakers with a practical tool for designing affordable, reliable renewable microgrids that support universal energy access objectives while advancing decarbonization goals.

Finally, the scalable nature of the framework positions this work as a foundation for broader adoption of sustainable off-grid energy solutions worldwide. \hl{Future research directions should explore the integration of complementary renewable sources (wind, small hydro), long-duration storage technologies (hydrogen, flow batteries), and demand response mechanisms to further enhance reliability and economic performance of off-grid microgrids.}

\section*{Declaration of generative AI use}

During the preparation of this work the authors used ChatGPT to assist with writing clarity and language polishing. After using this tool, the author reviewed and edited the content as needed and takes full responsibility for the content of the published article.

\section*{Acknowledgments}
This work was supported by the RACE for 2030 CRC
under Grant 21.PhD.N3.000006 and the Australian Research Council (ARC) Discovery Early Career Researcher Award (DECRA) under Grant DE230100046.

\section*{Declaration of competing interest}
The authors declare that they have no known competing financial interests or personal relationships that could have appeared to influence the work reported in this paper.

\section*{Data availability}
Data will be made available on request.

\bibliographystyle{elsarticle-num}
\bibliography{ref}

\clearpage
\onecolumn
\appendix
\appendix
\section{Robustness Analysis: Battery-Only Node Under Equipment Failure}
\label{appendix:node13}

To complement the node-level failure analyses presented in the main text (Figs.~8--10), this appendix examines the operation of Node~13, a battery-only node with no local PV installation -- under a battery outage scenario. 

\begin{figure}[htbp]
    \centering
    \includegraphics[width=\textwidth]{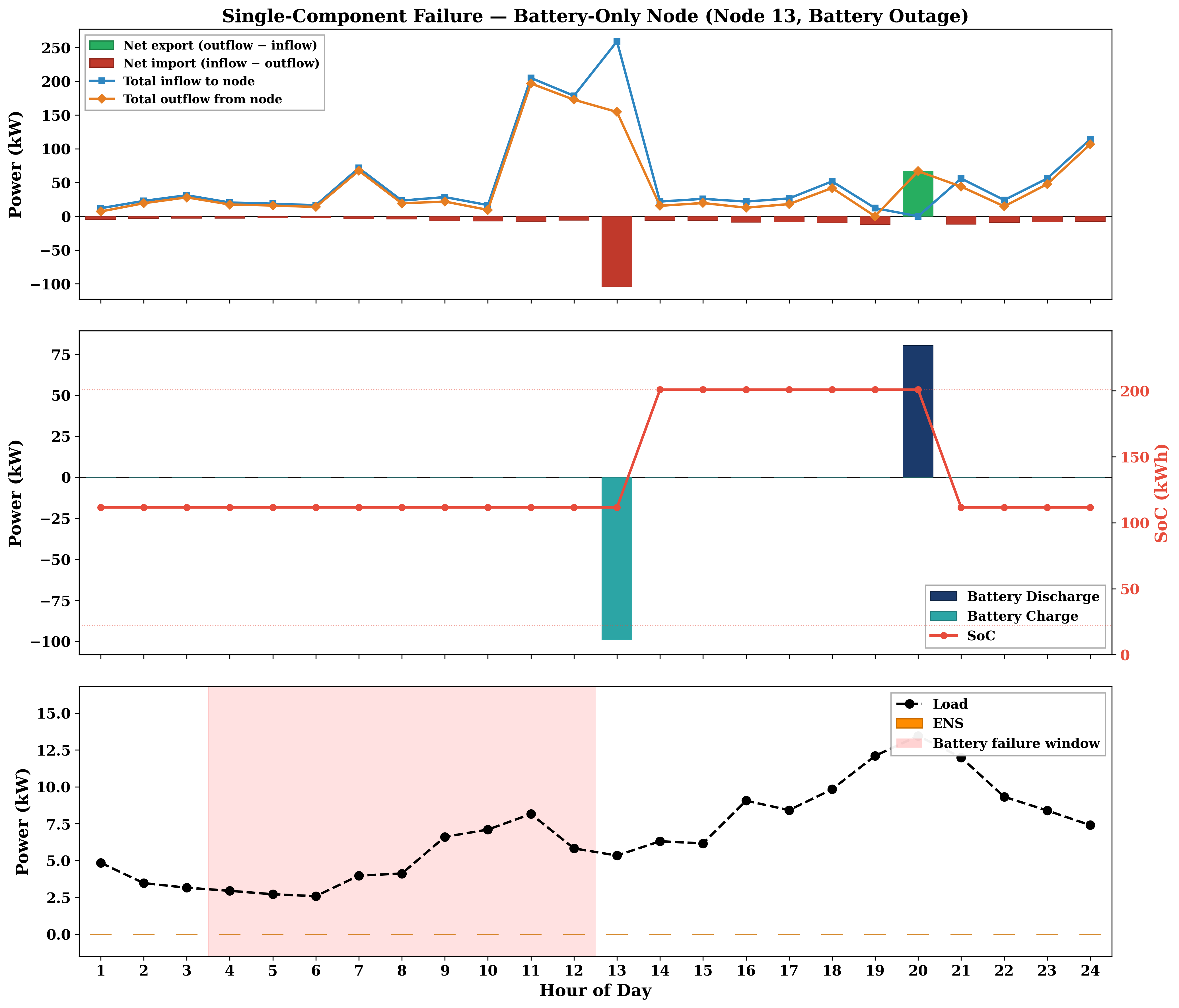}
    \caption{Hourly operational profile of Node~13 (a battery-only node with no local PV) under a single-component failure scenario (battery outage, hours~4--12): net power exchange with the network (top), battery charge/discharge and state of charge (middle), load demand and ENS (bottom). The shaded region indicates the battery failure window.}
    \label{Fig.11}
\end{figure}

Fig.~\ref{Fig.11} depicts the hourly operational profile of Node~13 under a 9-hour battery outage spanning hours~4--12. During the failure window, battery operation is fully suppressed -- charging and discharging are both zero, and the state of charge remains frozen at its pre-failure value of 111.6~kWh. Despite losing its only local asset during a period that spans both overnight base-load hours and the core morning demand ramp, Node~13 does not experience any supply disruption. The entire load during the outage is met exclusively through power imports from neighboring nodes, as confirmed by the net import values in the top panel closely tracking the load profile.

Once the failure clears at hour~13, the battery immediately charges at 99.2~kW to recover its state of charge, drawing energy from the network during peak solar hours when system-wide surplus is available. The battery subsequently discharges at hour~20 (80.4~kW), exporting stored energy to support the wider network during evening peak demand, before returning to its initial SoC of 111.6~kWh -- confirming the cyclic constraint (Eq.~33 in the main text). The expected energy not served remains zero throughout all 24~hours.

This result is particularly significant because it demonstrates that even a node with no local generation whatsoever can maintain uninterrupted supply during an extended equipment outage, relying solely on the redundancy and power-sharing capacity of the distributed network. The finding validates that the topology-aware planning approach provides inherent resilience not only to nodes with local generation backup but also to purely import-dependent nodes at the network periphery.

\end{document}